\documentclass[lettersize,journal]{IEEEtran}
\usepackage{amsmath,amsfonts}
\usepackage{algorithmic}
\usepackage{algorithm}
\usepackage{array}
\usepackage{textcomp}
\usepackage{stfloats}
\usepackage{url}
\usepackage{verbatim}
\usepackage{graphicx}
\usepackage{cite}
\usepackage{subfigure}
\usepackage{comment}
\usepackage{xcolor}
\usepackage{float}
\usepackage{booktabs}
\usepackage{multirow}
\usepackage{caption}
\usepackage[table,xcdraw]{xcolor}
\begin{document}

\title{LightRot: A Light-weighted Rotation Scheme and Architecture for Accurate Low-bit Large Language Model Inference}

\author{Sangjin Kim,~\IEEEmembership{Member,~IEEE,}
        Yuseon Choi,~\IEEEmembership{Graduate Student Member,~IEEE,}
        Jungjun Oh,~\IEEEmembership{Graduate Student Member,~IEEE,}
        Byeongcheol Kim,~\IEEEmembership{Graduate Student Member,~IEEE,}
        and Hoi-Jun Yoo,~\IEEEmembership{Fellow,~IEEE,}
        % <-this % stops a space
\thanks{Manuscript received 2 December 2024; revised 3 March 2025; accepted 28 March 2025. 

This work was supported in part by Samsung Electronics Company Ltd. under Grant IO201207-07799-01 and in part by the  Institute of Information and Communications Technology Planning and Evaluation (IITP) grant funded by the Korea Government (MSIT) (2022-0-01170, PIM Semiconductor Design Research Center) (Corresponding author: Hoi-Jun Yoo)

The Authors are with the PIM Semiconductor Design Research Center (AI-PIM), Daejeon 34141, South Korea (e-mail: sangjinkim@kaist.ac.kr; hjyoo@kaist.ac.kr)

\noindent\copyright~2025 IEEE. Personal use of this material is permitted. Permission from IEEE must be obtained for all other uses, in any current or future media, including reprinting/republishing this material for advertising or promotional purposes, creating new collective works, for resale or redistribution to servers or lists, or reuse of any copyrighted component of this work in other works.

S. Kim, Y. Choi, J. Oh, B. Kim and H. Yoo, ``LightRot: A Light-Weighted Rotation Scheme and Architecture for Accurate Low-Bit Large Language Model Inference,'' in IEEE Journal on Emerging and Selected Topics in Circuits and Systems, vol. 15, no. 2, pp. 231-243, June 2025, doi: 10.1109/JETCAS.2025.3558300.}}

% The paper headers
\markboth{IEEE Journal on Emerging and Selected Topics in Circuits and Systems, VOL. XX, No. X, XX, 2025}%
{Shell \MakeLowercase{\textit{et al.}}: A Sample Article Using IEEEtran.cls for IEEE Journals}

\maketitle

\begin{abstract}
As large language models (LLMs) continue to demonstrate exceptional capabilities across various domains, the challenge of achieving energy-efficient and accurate inference becomes increasingly critical. This work presents LightRot, a lightweight rotation scheme and dedicated hardware accelerator designed for low-bit LLM inference. The proposed architecture integrates Grouped Local Rotation (GLR) and Outlier Direction Aligning (ODA) algorithms with a hierarchical Fast Hadamard Transform (FHT)-based rotation unit to address key challenges in low-bit quantization, including the energy overhead of rotation operations. 
The proposed accelerator, implemented in a 28nm CMOS process, achieves a peak energy efficiency of 27.4 TOPS/W for 4-bit inference, surpassing prior state-of-the-art designs. Unlike conventional approaches that rely on higher-precision inference or evaluate on basic language modeling tasks like GPT-2, LightRot is optimized for advanced models such as LLaMA2-13B and LLaMA3-8B. Its performance is further validated on MT-Bench, demonstrating robust applicability to real-world conversational scenarios and redefining benchmarks for chat-based AI systems. By synergizing algorithmic innovations and hardware efficiency, this work sets a new paradigm for scalable, low-bit LLM inference, paving the way for sustainable AI advancements.

\end{abstract}


\section{Introduction}
\IEEEPARstart{N}{owadays}, large language models (LLMs) have demonstrated remarkable capabilities across a wide range of tasks, including language translation, code generation, and conversational AI. Their profound impact on industries and daily interactions has solidified their role as a cornerstone of modern artificial intelligence \cite{llm_intro_ref1, llama2, llama3, mistral}. However, the immense size of LLMs presents significant challenges, particularly in terms of memory and computational demands. This necessitates the development of lightweight techniques such as quantization to optimize their performance. These optimizations are critical not only for server-side inference, where large-scale computational resources are available, but also for on-device applications, where resources are constrained and smaller LLMs must perform efficiently.

Quantization has emerged as a key technique to address the computational and memory challenges posed by LLMs, focusing on reducing the precision of data representations while maintaining acceptable accuracy. As illustrated in Fig. 1 (a), the quantization process for LLMs involves three main data types: weights (W), input activations (IA), and key-value caches (KV). Among these, weights are typically quantized after training through post-training quantization, allowing them to be stored in a compressed form without impacting runtime performance. In contrast, input activations and key-value caches are generated dynamically during inference, requiring efficient online quantization methods to minimize runtime overhead.

A significant challenge lies in the distribution of input activations, which often vary significantly across channels and exhibit large outliers. These heavy-tailed distributions lead to suboptimal quantization performance, as conventional quantization methods struggle to capture the variations without introducing substantial errors. Moreover, during the generation stage, frequent access to weight parameters stored in external memory creates a critical bottleneck, both in terms of energy consumption and inference speed. While weight-only quantization \cite{awq, owq, squeezellm} addresses these issues to some extent by focusing on the weight data, it does not fully resolve the challenges associated with input activations and key-value caches, which remain major contributors to computational and memory costs.

\begin{figure}[]
\centering
\subfigure[]{
\includegraphics[width=3.4in]{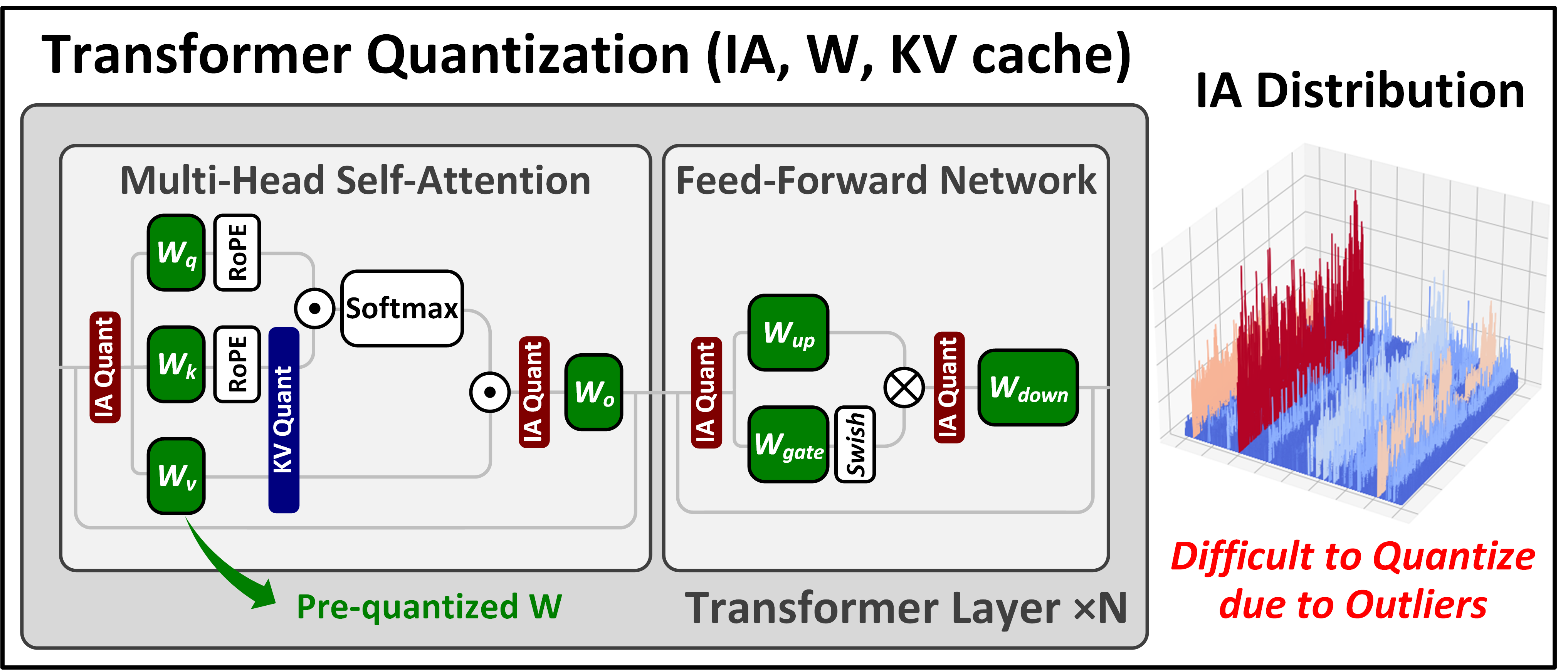}
\centering
}
\subfigure[]{
\includegraphics[width=3.4in]{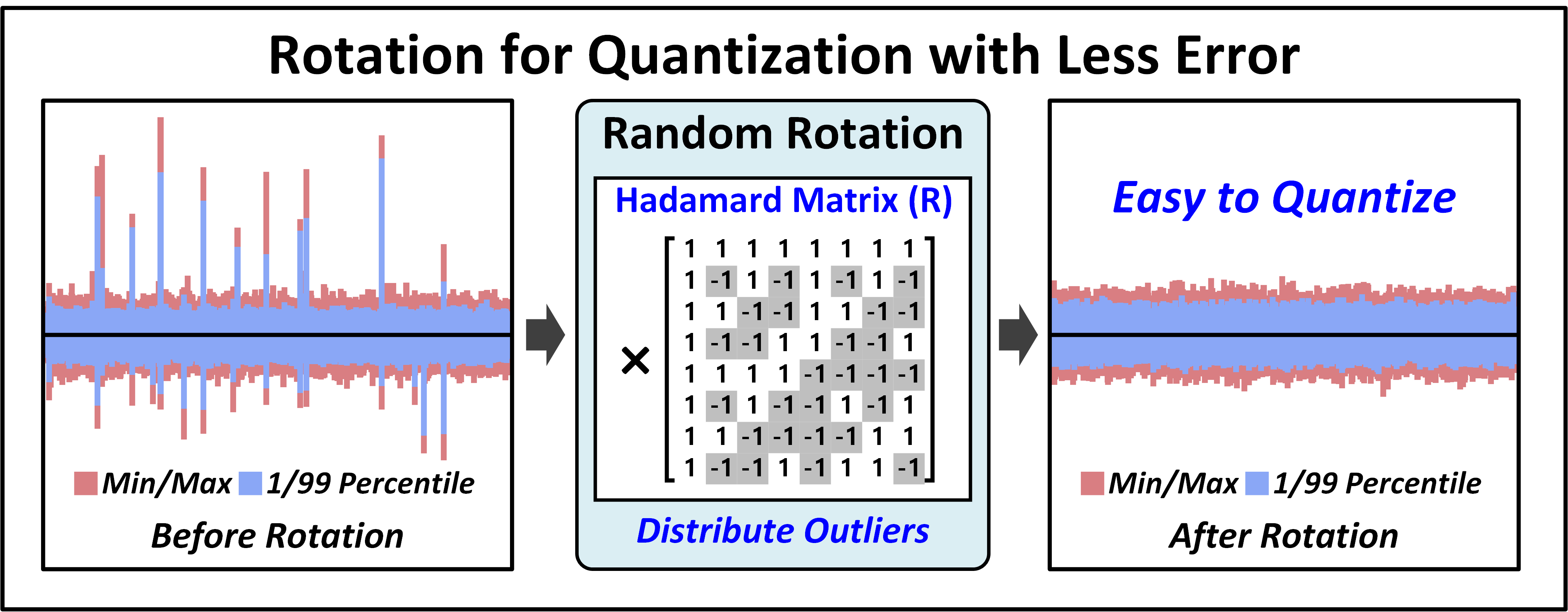}
\centering
}
\centering
\caption{
(a) Challenge of LLM Quantization by Outlier and (b) Concept of Rotation
}
\label{fig_11}
\vspace{-6pt}
\end{figure}

As shown in Fig. 1 (b), rotation-based quantization has emerged as a promising solution to address the limitations of conventional quantization methods. By applying a rotation matrix to tensors, such as W, IA, and KV, this approach reduces the impact of outliers and improves quantizability. Specifically, the rotation operation reorients the data distribution, minimizing the variance in certain directions and enabling a more even representation suitable for quantization.

One of the key principles behind this method is the computational invariance property of rotation, which ensures that tensor rotations do not affect the overall network outputs when paired with proper inverse rotations after computation. This property enables more flexible manipulation of tensor representations without compromising model performance. Unlike weight-only quantization, rotation-based quantization effectively extends its benefits to IA and KV, which are significant contributors to the memory and computational costs of LLMs. As a result, this method achieves a more comprehensive reduction in both memory footprint and computational overhead, making it a compelling alternative for low-bit inference scenarios \cite{spinquant, quarot, flatquant}.

In this work, we propose LightRot, a lightweight rotation-based quantization scheme designed to optimize low-bit inference for LLMs. LightRot significantly reduces the computational burden by lowering the algorithmic complexity of the rotation process, making it more practical for real-time and resource-constrained environments. Additionally, the proposed method enhances the accuracy of rotation-based quantization by refining the rotation operation, ensuring minimal distortion of the quantized tensors.

To complement the algorithm, we also introduce a dedicated hardware architecture that accelerates the rotation process. This architecture is optimized to handle the computational demands of rotation-based quantization efficiently, while maintaining a low area and power footprint. By addressing both the algorithmic and hardware challenges, LightRot achieves substantial improvements in memory efficiency, computational cost, and overall inference performance, making it a comprehensive solution for low-bit LLM applications.

\section{Background and Challenges}

\begin{figure*}[]
\includegraphics[width=6.0in]{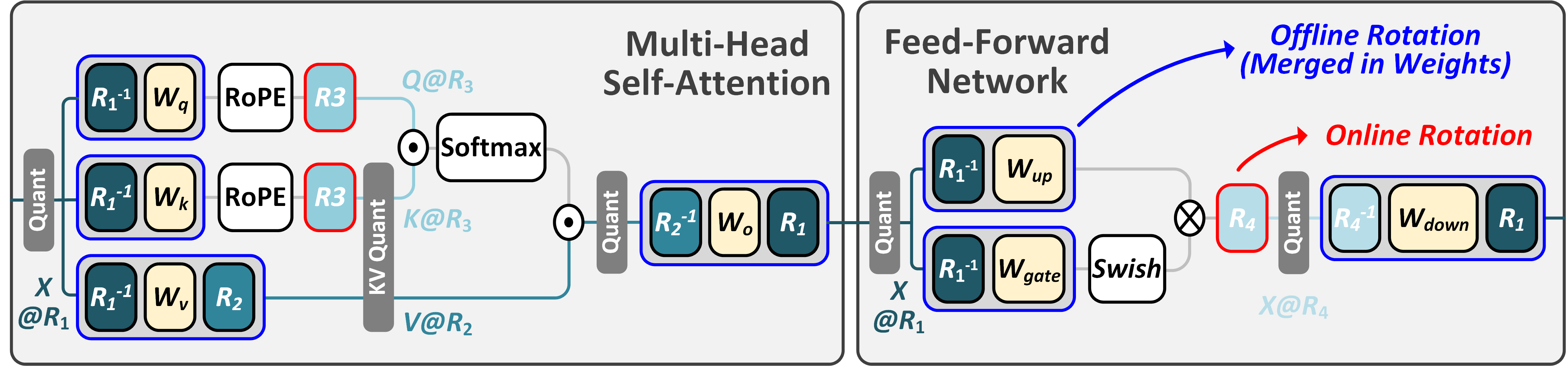}
\centering
\caption{Applying Rotation in LLM Model with Online/offline Rotation.}
\label{fig_2}
\vspace{-0.2cm}
\end{figure*}

One of the most challenging aspects of quantizing LLMs lies in the distribution of input activations (IA). The IA distribution often varies significantly across channels, with some channels exhibiting notably higher magnitudes compared to others, forming what are referred to as outlier channels. These outlier channels play a crucial role in determining the overall accuracy of the model, as their improper handling can lead to significant performance degradation. Interestingly, the magnitudes of these outlier channels remain relatively stable across tokens and batches, making them a consistent characteristic of IA distributions. This stability can be effectively leveraged during the quantization process to optimize performance.

Addressing these unique characteristics, prior studies on low-bit LLM inference, such as \cite{awq, smoothquant, owq, atom, quarot, spinquant}, have focused on developing methods to accurately and efficiently quantize IAs or weights (W) while minimizing accuracy loss. These works have advanced the field by incorporating techniques that account for IA distribution properties, particularly outlier channels, which present both challenges and opportunities for improved quantization.

\subsection{Channel scale-aware Quantization}
A straightforward approach to addressing the challenges of IA quantization is to adjust the scaling factors for each channel based on their saliency. Some methods \cite{smoothquant, awq} adopt this strategy by applying different scaling factors to each channel according to its saliency, making the distributions across channels more even. For example, one approach \cite{smoothquant} leverages the observation that weights (W) with flat distributions are easier to quantize than activations (IA). By migrating the quantization difficulty from activations to weights, these methods smooth IA outliers through offline scaling, making them more quantizable. Another approach \cite{awq} applies a similar strategy during the generation stage, reducing weight precision while maintaining IA in floating-point form to alleviate bottlenecks caused by weight-efficient matrix accumulation.

However, while scaling factor methods ensure the range of salient channels is preserved, they inherently limit the resolution of the quantized data due to the inherent limitations of scaling. This trade-off imposes constraints on their quantization effectiveness. To address this, mixed precision strategies \cite{atom, owq}, assign higher precision to salient channels and lower precision to less critical ones. This approach achieves a balance between maintaining accuracy and lowering effective precision. However, it introduces additional challenges, such as latency overhead on GPUs and multi-precision logic overhead on NPUs, posing challenges in practical implementations.

\subsection{Rotation-based Quantization}
Rotation-based quantization offers an alternative approach to addressing the challenges posed by IA outliers by transforming the data distribution through matrix rotation. As shown in Fig. 1 (b), a rotation matrix (e.g., Random Hadamard matrix) is multiplied to the IA matrix, effectively redistributing outliers across channels and flattening the overall distribution. This operation minimizes the directional variance across channels, making the IA data more quantizable. One key advantage of this approach is the computational invariance guaranteed by the rotational property of the Hadamard matrix. By applying the inverse matrix to the weights (W), the overall computational correctness of the network is preserved, as proved in \cite{quarot}.

In practical implementations, rotation matrices can be categorized into offline and online types, depending on whether the rotation can be pre-merged with the weights. As illustrated in Fig. 2, transformer layers typically involve four rotation matrices (R1–R4). Among these, R1 and R2 can be pre-merged with the weights during offline processing, reducing runtime overhead. However, R3 and R4 require online rotation, as they cannot be merged with the weights. Fortunately, Hadamard matrices consist only of ±1 values, enabling online rotation through simple addition operations on IA, without the need for complex computations.

Some advanced methods, such as \cite{spinquant}, further optimize rotation-based quantization by employing learned rotation matrices instead of random Hadamard matrices. While this distribution-aware approach improves quantization performance, it significantly increases computational complexity and restricts its applicability to offline rotations. In contrast, the proposed LightRot introduces a lightweight, distribution-aware rotation algorithm that maintains low computational complexity and extends its applicability to online rotations, providing a more comprehensive and practical solution for both offline and online rotations in low-bit LLM inference.

\begin{figure}[]
\includegraphics[width=3.4in]{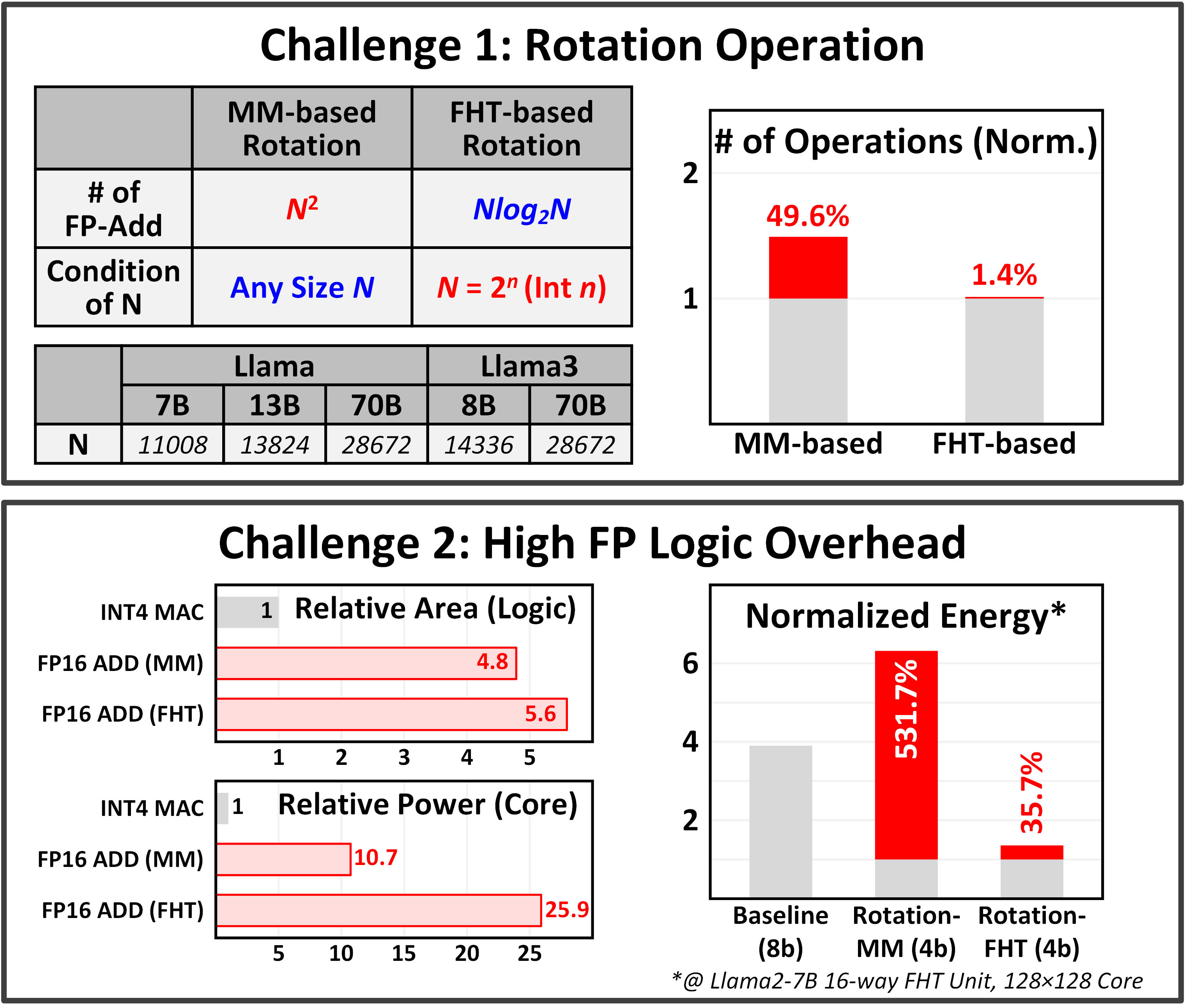}
\centering
\caption{Challenges of Rotation-based Quantization}
\label{fig_2_2}
\vspace{-0.2cm}
\end{figure}

\subsection{Challenges of Rotation-based Quantization Hardware}
Despite its advantages, implementing rotation-based quantization in hardware introduces significant challenges, as summarized in Fig. 3. These challenges stem primarily from the design of the rotation unit and the high cost of floating-point (FP) operations required during rotation.

The first challenge is related to the datapath design of the rotation unit. To process online Hadamard matrices ($R_3$ and $R_4$), a rotation unit is essential. Two primary approaches are commonly used: matrix multiplication (MM)-based rotation and fast Hadamard transform (FHT)-based rotation. The MM-based approach multiplies the Hadamard matrix, consisting only of $+1$ and $-1$, with the IA tensor. While this requires only addition operations, it involves $\mathcal{O}(N^2)$ computations for a vector of length $N$. On the other hand, FHT-based rotation \cite{quip} employs a divide-and-conquer recursive algorithm, significantly reducing the computational complexity to $\mathcal{O}(N \log_2 N)$.

However, FHT-based rotation can only be applied to dimensions that are powers of two, whereas many commercial LLM architectures use non-power-of-two dimensions to optimize memory and computational efficiency. In these cases, designers are forced to either rely solely on MM-based rotation or combine MM-based and FHT-based approaches. Solely using MM-based rotation results in high computational overhead, with rotation operations accounting for up to 49.6\% of the multiply-accumulate (MAC) workload. Alternatively, combining MM-based and FHT-based approaches introduces a substantial hardware cost.

The second challenge stems from the high cost of FP operations in the rotation unit. While rotation-based quantization allows layer computations within the NPU core to use efficient INT operations, the rotation unit itself operates on IA before quantization, requiring costly FP operations. FP units consume significantly more resources, with power and area costs that are $25.9\times$ and $5.6\times$ higher, respectively, compared to INT4 MAC units. Due to the high logic cost, even though FHT-based rotation reduces the number of operations compared to INT MAC, it still incurs significant energy consumption. This results in over 35.7\% energy overhead in the layer configuration of the LLaMA2-7B model, making it a bottleneck for energy-efficient LLM inference.

\section{Proposed LightRot Algorithm}

\begin{figure*}[]
\includegraphics[width=6.0in]{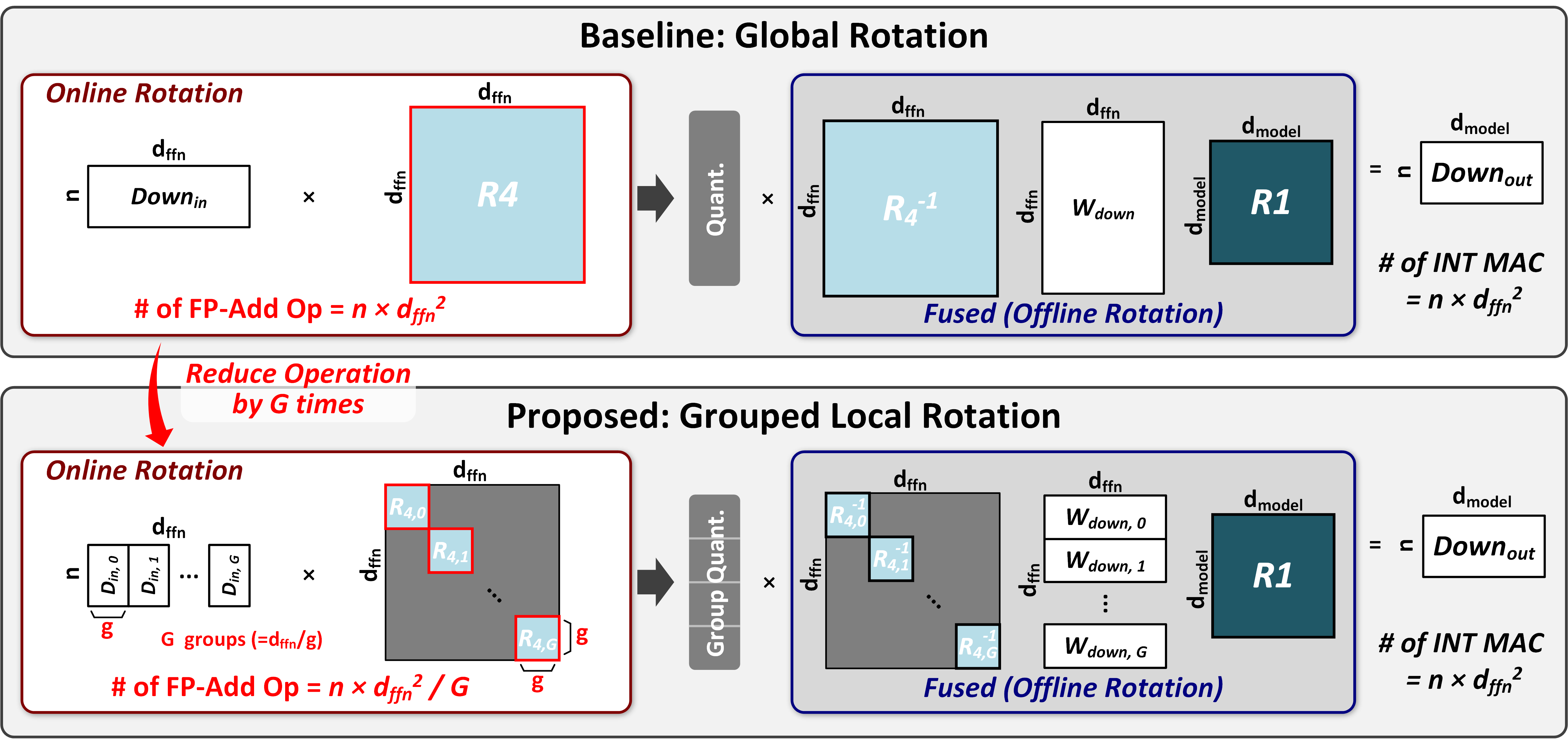}
\centering
\caption{Concept of Grouped Local Rotation.}
\label{fig_2_3}
\vspace{-0.2cm}
\end{figure*}

In this section, we introduce the LightRot algorithm, designed to support low-cost and high-accuracy low-bit quantized LLM inference. LightRot builds upon the existing rotation algorithm \cite{quarot} to maintain the performance of low-bit quantized LLMs while simultaneously reducing rotation costs through Grouped Local Rotation (GLR) and improving rotation accuracy with Outlier Direction Aligning (ODA). GLR reduces hardware costs by dividing rotation operations into smaller groups and applying localized rotation within each group. Meanwhile, ODA leverages the consistent outlier distribution characteristics of LLMs and the properties of Hadamard-based rotation to make post-rotation distributions more quantizable, minimizing accuracy loss during quantization.

\subsection{Grouped Local Rotation}
\begin{comment}
\begin{figure}[]
\includegraphics[width=3.4in]{Figure/F4.png}
\centering
\caption{Motivation of Grouped Local Rotation.}
\label{fig_2_4}
\vspace{-0.2cm}
\end{figure}
\end{comment}

\begin{figure}[]
\includegraphics[width=3.4in]{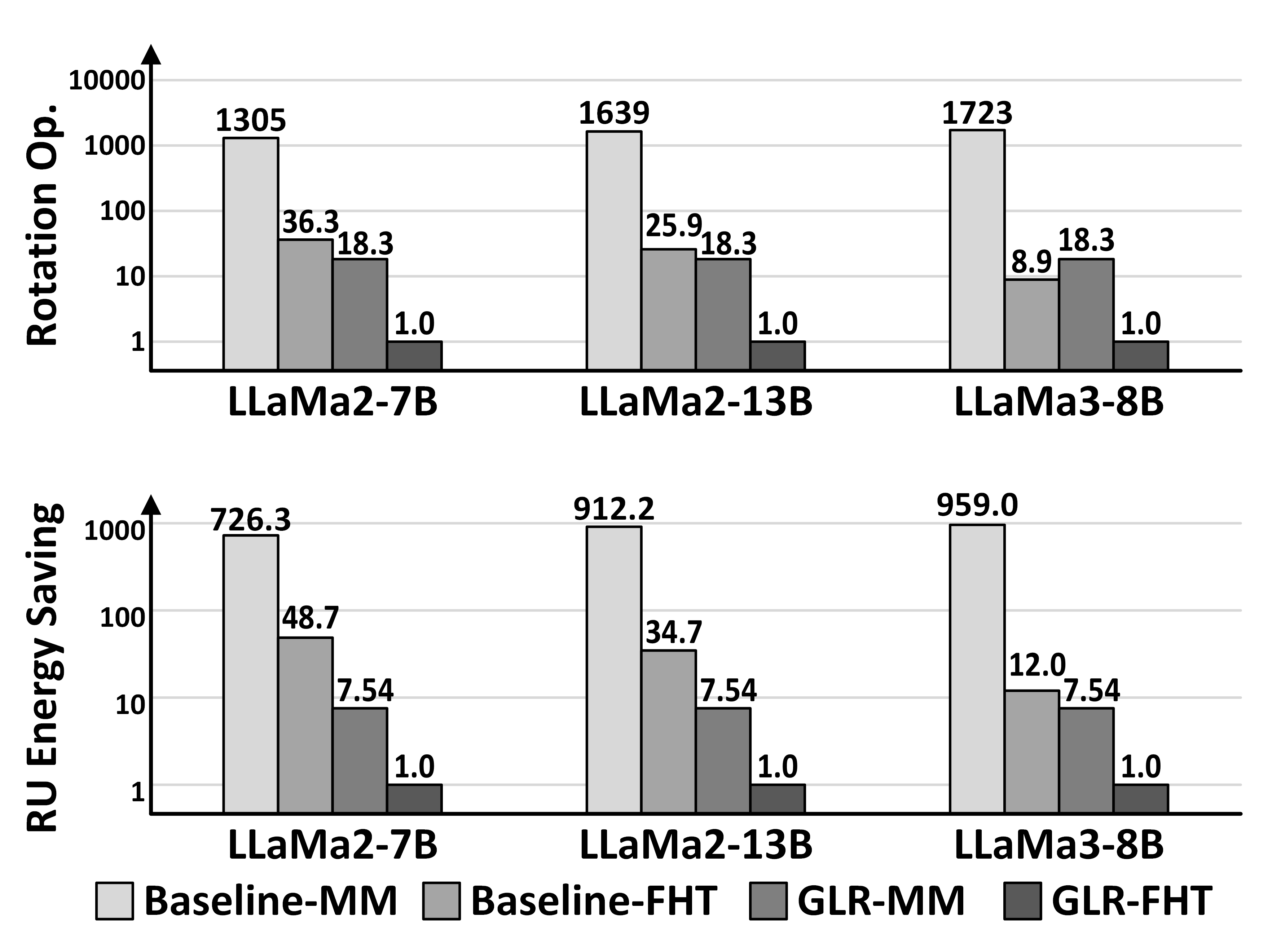}
\centering
\caption{Rotation Operation and Rotation Unit Energy Reduction by GLR.}
\label{fig_2_5}
\vspace{-0.2cm}
\end{figure}

\begin{figure}[]
\includegraphics[width=3.4in]{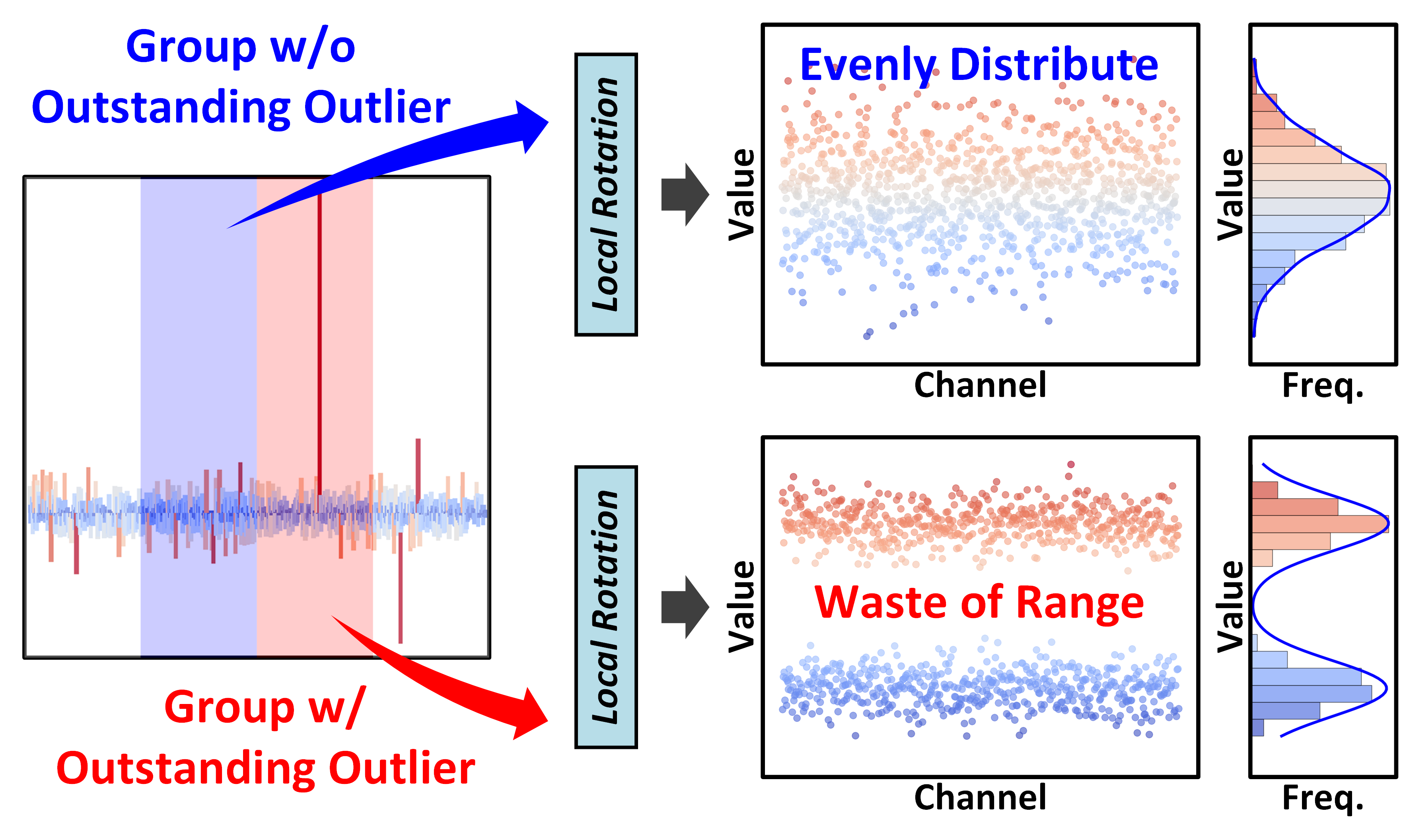}
\centering
\caption{Motivation of Outlier Direction Aligning.}
\label{fig_2_6}
\vspace{-0.2cm}
\end{figure}
Quantized LLMs operate with quantized input activations (IA) and weights (W) in each layer. The multiplication of IA and W is performed in integer precision, enabling efficient integer accumulation across thousands or tens of thousands of channel dimensions commonly found in LLM architectures \cite{llama2, llama3}. However, to maintain accuracy, group quantization is often applied, requiring FP16 precision scale factors and biases \cite{quarot}. Consequently, the output activation (OA) is computed in FP precision and quantized for use as the IA in subsequent layers.

For layers requiring online rotation, Hadamard matrices are multiplied with IA before quantization. Despite Hadamard matrices being composed solely of $+1$ and $-1$, enabling addition/subtraction-based computation, this operation must still be performed in FP precision, incurring significant computational costs. As illustrated in Fig. 4, online rotations involving $R_4$ matrices (e.g., between the first and second feed-forward network (FFN) layers in transformer architectures) introduce substantial computational overhead. In many LLM models, gated linear units (GLU) or their variants, such as SwiGLU \cite{swiglu}, are used for accuracy enhancement between the two FFN layers. Since activation rotations between these layers cannot be merged into weights, the number of operations for $R_4$ becomes $n \times d_\text{ffn}^2$, where $n$ is the number of tokens and $d_\text{ffn}$ is the FFN dimension. This results in $n \times d_\text{ffn}^2$ FP-Add operations, making the reduction of rotation costs crucial.

GLR addresses this issue by dividing the IA into multiple groups and applying smaller rotation matrices to each group instead of using a single large rotation matrix. The principle of rotation-based quantization relies on redistributing outliers across channels to enhance quantizability. GLR achieves sufficient quantizability by dispersing outliers locally within each group rather than across all channels. To implement this, GLR partitions the IA into $G$ groups, each consisting of $g$ channels, where $G = d_\text{ffn} / g$. For each group, a $g \times g$ rotation matrix is applied, resulting in a total of $G$ local rotations. Consequently, the total number of FP-Add operations is reduced to $n \times d_\text{ffn}^2 / G$.

Fig. 5 summarizes the reduction in rotation operations and energy achieved by GLR. Both matrix multiplication (MM)-based and fast Hadamard transform (FHT)-based rotation units are considered, following the discussion in Section II-C. For non-power-of-two channel dimensions, FHT-based units require the integration of MM-based units, leading to inefficiencies. GLR mitigates this issue by using $g \times g$ rotation matrices, allowing $g$ to be chosen as a power-of-two dimension, even if $d_\text{ffn}$ is not. This enables FHT-only implementations. In our benchmark evaluations, $g$ was set to 128 to match with the group sizes for group quantizations in previous works \cite{atom, quarot}, and GLR was tested on LLaMA2-7B/13B and LLaMA3-8B models. With MM-based rotation units, GLR reduced operations by 71–94$\times$ and energy by 96–127$\times$. With FHT-based rotation units, GLR reduced operations by 9–36$\times$ and energy by 12–49$\times$.

\subsection{Outlier Direction Aligning}

\begin{figure*}[]
\includegraphics[width=6.4in]{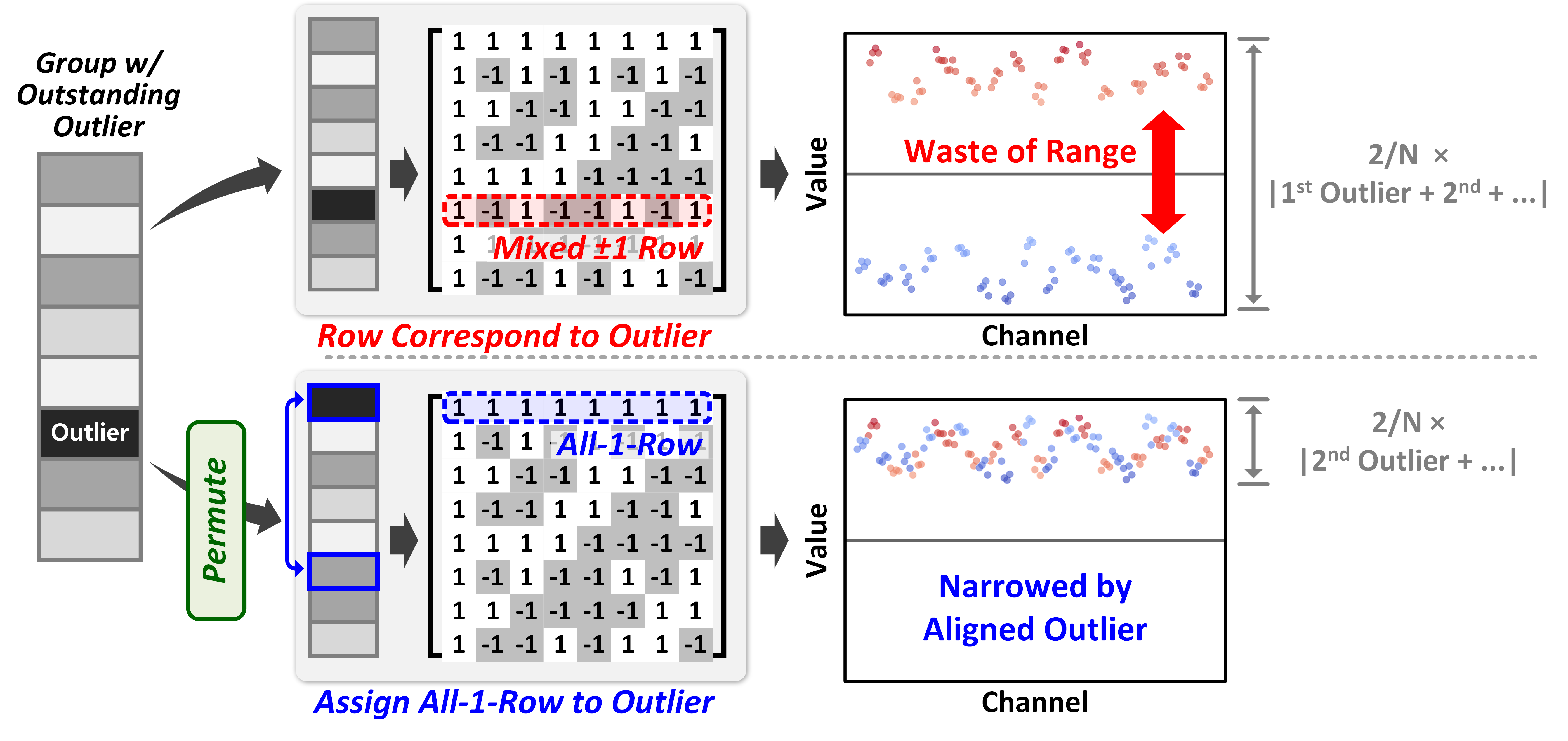}
\centering
\caption{Concept of Outlier Direction Aligning.}
\label{fig_2_7}
\vspace{-0.2cm}
\end{figure*}

\begin{figure}[]
\includegraphics[width=3.4in]{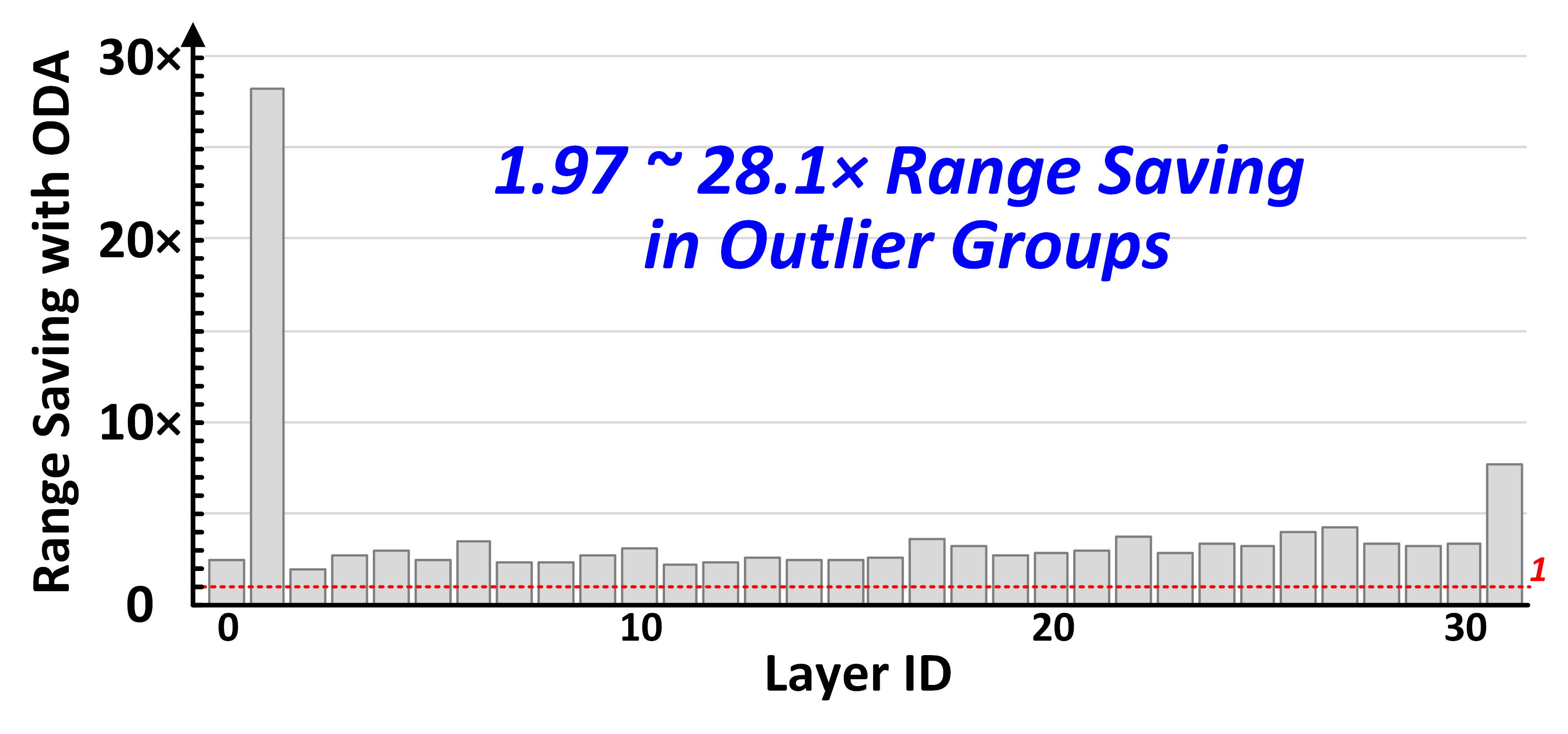}
\centering
\caption{Range Reduction with the ODA. Evaluated on the top 5\% Outlier Group.}
\label{fig_2_8}
\vspace{-0.2cm}
\end{figure}

Fig. 6 illustrates the data distribution after applying GLR, providing motivation for the Outlier Direction Aligning (ODA) algorithm. IA quantization is particularly challenging due to the presence of outliers, which rotation operations can mitigate. However, the impact of outliers in practice is heavily influenced by their magnitude. While most outliers fall within 100 times the inlier data range and can be easily dispersed by rotation, certain extreme cases, termed {\it{\textbf{outstanding outliers}}} exhibit magnitudes thousands to tens of thousands of times larger than the inlier range, making them difficult to disperse effectively. These outstanding outliers significantly affect post-rotation distributions.

Groups without outstanding outliers achieve near-normal distributions after rotation, enabling improved quantizability and accurate quantization. In contrast, groups containing outstanding outliers exhibit uneven distributions even after rotation. Due to the large magnitude of these outliers, they dominate the rotated data, creating bimodal distributions where the range is wasted, leading to significant quantizabilty drops during quantization.

Fig. 7 illustrates the principle behind ODA, which addresses this issue by leveraging the properties of Hadamard matrices, commonly constructed using Sylvester construction \cite{quarot}. Sylvester construction is a recursive method for generating Hadamard matrices while maintaining orthogonality and compatibility with FHT algorithms. Starting from the base matrix:
\[
H_1 = 
\begin{bmatrix}
1 & 1 \\
1 & -1
\end{bmatrix},
\]
larger Hadamard matrices are constructed recursively as:
\[
H_{n+1} = 
\begin{bmatrix}
H_n & H_n \\
H_n & -H_n
\end{bmatrix}.
\]
This method ensures that each row alternates between $+1$ and $-1$, except for the first row, which consists entirely of $+1$ values (the "all-1-row").

In rotation operations, outlier channels are multiplied by the corresponding row of the Hadamard matrix. Outstanding outliers result in wasteful range utilization, even when dispersed across channels. To address this, ODA aligns the largest outlier in each group with the all-1-row of the Hadamard matrix before performing rotation. By doing so, the channel containing the outstanding outlier is multiplied solely by $1$, while other channels are rotated normally using $+1$ and $-1$. This prevents bimodal distributions and ensures a more gaussian, quantizable distribution after rotation.

However, while this approach effectively narrows the range, it introduces a biased distribution, necessitating asymmetric quantization, as expressed in the following equations:
\[
q_{\text{IA}} = \text{round}\left(\frac{\text{IA} - \text{IA}_{\text{bias}}}{\text{IA}_{\text{scale}}}\right)
\]

\[
\text{OA} = \frac{1}{\text{IA}_{\text{scale}}} \cdot \left( (\text{IA} \cdot w) - (\text{IA}_{\text{bias}} \cdot w) \right)
\]

Here, $\text{IA}_{\text{scale}}$ and $\text{IA}_{\text{bias}}$ are shared within the group, allowing the second term, $\text{IA}_{\text{bias}} \cdot w$, to become a constant multiplied by the weights. Consequently, the channel-wise sum of $w$ can be precomputed offline at the group level, significantly reducing the computation cost.

Furthermore, ODA requires a permutation step to align the weights prior to rotation. However, as prior studies \cite{smoothquant, awq, atom, spinquant} have demonstrated, outlier indices remain consistent across batches, allowing the permutation order to be pre-determined using a calibration set, such as WikiText-2 perplexity benchmarks \cite{wikitext2}.

Fig. 8 summarizes the range savings achieved by ODA in the top 5\% outlier groups for the LLaMA2-7B model on the WikiText-2 benchmark. These groups are highly critical to accuracy due to their outstanding outliers. ODA effectively reduces the range in these groups, achieving a minimum 1.97$\times$ and maximum 28.1$\times$ reduction, with an average savings of 3.87$\times$. The first layer showed the most significant range reduction, attributed to the higher prevalence of outstanding outliers compared to other layers. This reduced range eliminates wasted capacity during quantization, enhancing quantizability.

\section{Algorithm Experiments}

\begin{table}[]
\caption{Perplexity Comparison on WikiText-2 Dataset}
\begin{tabular}{@{}lccccc@{}}
\toprule
\multicolumn{1}{c}{\multirow{2}{*}{\textbf{PPL↓}}} & \multicolumn{1}{l}{\multirow{2}{*}{\textbf{Group Size}}} & \multicolumn{2}{c}{\textbf{LLaMa-2}} & \multicolumn{1}{l}{} & \textbf{LLaMa-3} \\ \cmidrule(lr){3-4} \cmidrule(l){6-6} 
\multicolumn{1}{c}{}                               & \multicolumn{1}{l}{}                                     & 7B                & 13B              &                      & 8B               \\ \midrule
FP16                                               & -                                                        & 5.47              & 4.88             &                      & 6.14             \\ \midrule
SmoothQuant\cite{smoothquant}                                        & -                                                        & 83.12             & 35.88            &                      & -                \\
Quarot\cite{quarot}                                             & -                                                        & 6.10              & 5.40             &                      & -                \\
Quarot\cite{quarot}                                             & 128                                                      & 5.93              & 5.26             &                      & -                \\
SpinQuant\cite{spinquant}                                          & -                                                        & 5.96              & 5.24             &                      & 7.39             \\ \midrule
Rot Only                                           & -                                                        & 6.11              & 5.38             &                      & 7.93             \\
Rot+GLR                                            & -                                                        & 7.24              & 7.71             &                      & 11.13            \\
Rot+GLR+ODA                                        & -                                                        & 6.74              & 5.73             &                      & 10.9            \\ 
Rot+GQ                                             & 128                                                      & 5.87              & 5.22             &                      & 7.30             \\ \rowcolor[HTML]{E0F2F5}
Rot+GQ+GLR                                         & 128                                                      & 5.80              & 5.14             &                      & 7.12             \\ \rowcolor[HTML]{E0F2F5}
Rot+GQ+GLR+ODA                                     & 128                                                      & \textbf{5.73}     & \textbf{5.08}    & \textbf{}            & \textbf{6.98}    \\ \bottomrule 
\multicolumn{6}{r}{INT4 for all IA, W, KV} \\
\multicolumn{6}{r}{INT4-INT4 opeartion for all $Q, K, V$, Projection, FFN} \\
\multicolumn{6}{r}{FP16-INT4 operation for $Q K^T$, $SV$} \\
\end{tabular}
\end{table}

\begin{figure}[]
\centering
\subfigure[]{
\includegraphics[width=3.4in]{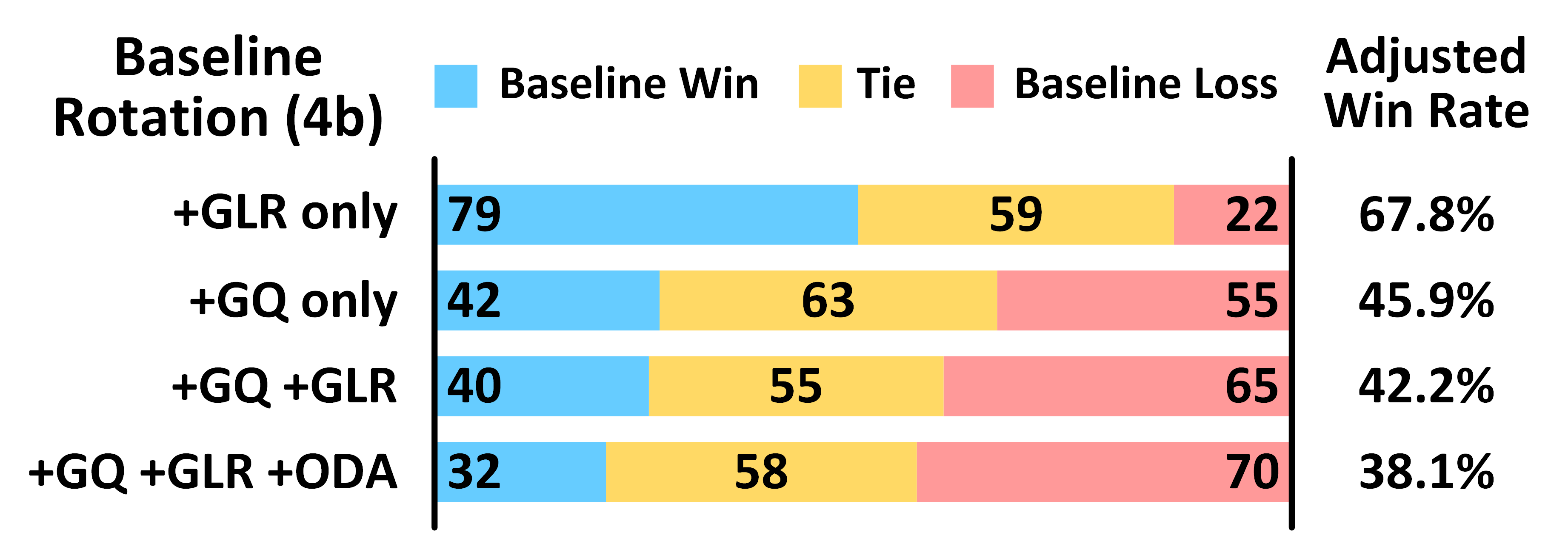}
\centering
}
\subfigure[]{
\includegraphics[width=3.4in]{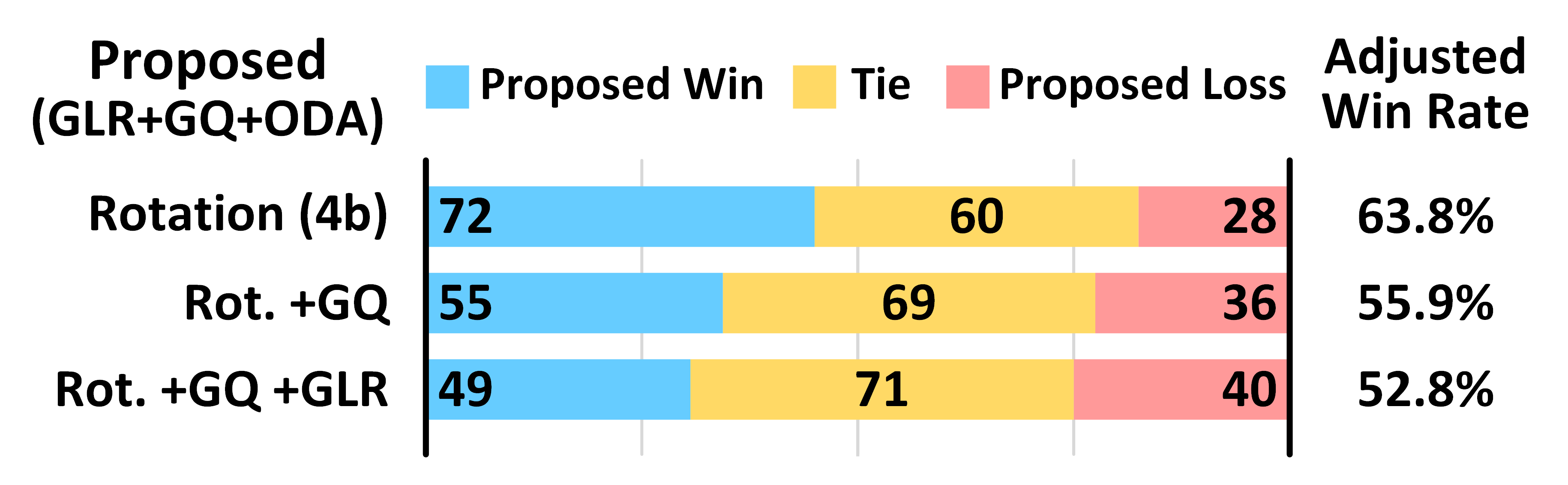}
\centering
}
\subfigure[]{
\includegraphics[width=3.4in]{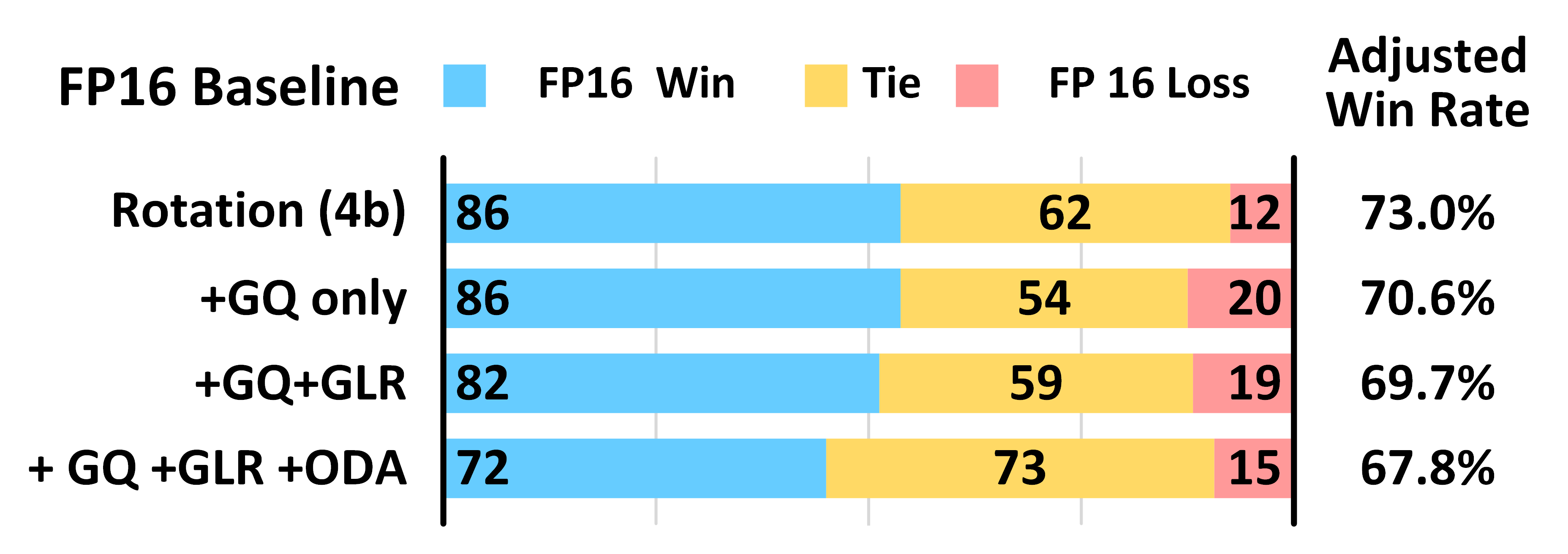}
\centering
}
\centering
\caption{Evaluation with LLM-as-a-Judge [17]. Win/Loss Rate against (a) Baseline Rotation Model (b) Proposed Model (c) FP16 Baseline Model}
\vspace{-6pt}
\end{figure}

In this section, we evaluate the performance of the proposed LightRot algorithm using multiple benchmarks and scenarios. The experiments demonstrate how the combined features—Group Quantization (GQ), Grouped Local Rotation (GLR), and Outlier Direction Aligning (ODA)—enhance performance over existing baseline methods, particularly in low-bit LLM inference tasks.

\subsection{Perplexity Evaluation on WikiText-2}
Table 1 presents the perplexity (PPL) results on the WikiText-2 dataset \cite{wikitext2} for various configurations, demonstrating the effectiveness of the proposed features (+GLR and +ODA) in improving quantization quality while maintaining low-bit inference efficiency.

The baseline rotation-only method achieved competitive PPL values compared to existing approaches like SpinQuant and Quarot. Also, using GLR and ODA together recovers some accuracy, but without GQ, it still shows lower accuracy. However, the addition of GLR and ODA demonstrates clear improvements in quantization performance. GLR significantly reduces computational costs while maintaining accuracy, and ODA further enhances performance by addressing the challenges posed by outstanding outliers.

When GLR is applied alone, there is a noticeable drop in accuracy due to the lack of scale alignment between groups. However, when GQ and GLR are used together with the same group size, the accuracy improves significantly. This improvement can be attributed to GQ's ability to compensate for the group-wise scale variations introduced by GLR. The synergy between GQ and GLR ensures consistent quantization quality across groups, highlighting the importance of jointly applying these techniques for optimal performance.

To ensure a fair comparison, Quarot was also evaluated with GQ applied. While Quarot demonstrated a minor PPL gap of 0.17 when group quantization was used, the proposed GLR exhibited a much larger improvement, reducing the PPL gap by 1.44. This stark contrast underscores the effectiveness of GLR in addressing group-level challenges during rotation and highlights its superiority over baseline methods.

Integrating all proposed features, including ODA, achieves the best perplexity results across all configurations. The final PPL values for the proposed method (+GQ+GLR+ODA) were 5.73 for LLaMA-2 (7B), 5.08 for LLaMA-2 (13B), and 6.98 for LLaMA-3 (8B). These results confirm that the proposed LightRot algorithm achieves superior quantization quality and inference performance compared to both baseline and state-of-the-art methods.

\subsection{Evaluation with LLM-as-a-Judge}
Many existing quantization studies \cite{quarot, spinquant} focus on evaluating accuracy using limited metrics such as perplexity or simple answering tasks. These evaluations often involve short output token lengths, making it difficult to assess the impact of KV cache quantization on long-form outputs. To address these limitations, we employ MT-Bench \cite{mtbench} to evaluate model accuracy in realistic chat scenarios using win-loss evaluations. This approach provides a more comprehensive understanding of how quantization affects model performance during extended interactions.

For this evaluation, we use conversational models (LLaMA3-8B-instruct). By employing chat-optimized models, the evaluation captures the interplay between quantization techniques and model behavior in real-world applications.

As shown in Fig. 9 (a), we compare the proposed methods against the baseline rotation method (Quarot) using win/lose rates. The proposed features (+GQ, +GQ+GLR, +GQ+GLR+ODA) are incrementally added, and their impact on performance is observed. With each added feature, the baseline's win rate consistently decreases, demonstrating the effectiveness of the proposed techniques. Notably, when all features are applied (+GQ+GLR+ODA), the baseline's adjusted win rate drops to 32.2\%, indicating clear superiority of the proposed approach.

Similar to the analysis in the perplexity evaluation, applying GLR alone leads to a reduction in win rate due to scale misalignment between groups. However, when GQ and GLR are applied together, the win rate improves significantly. This is because GQ compensates for group-wise scale variations introduced by GLR, ensuring a more consistent quantization quality across groups. This synergy highlights the importance of combining these techniques to achieve optimal performance.

Fig. 9 (b) presents a comparison between the fully-featured LightRot configuration (+GQ+GLR+ODA) and all other configurations, including baseline and alternative quantization methods. The proposed method maintains an adjusted win rate exceeding 50\% against all competitors, confirming its robustness and effectiveness. This consistency demonstrates that LightRot not only handles standard quantization challenges but also excels in more demanding evaluation settings.

Fig. 9 (c) compares the proposed methods against the FP16 baseline. While FP16 naturally achieves the highest win rate, the proposed LightRot configuration (+GQ+GLR+ODA) remains relatively competitive compared to other quantized configurations, demonstrating its ability to maintain strong accuracy even under low-bit precision.

By leveraging MT-Bench and LLM-as-a-Judge evaluation frameworks, this study highlights the practical advantages of the proposed algorithm in real-world conversational scenarios, where both short and extended outputs are critical.

\section{Proposed Accelerator}

To validate the performance of the proposed LightRot algorithm, we designed a dedicated hardware accelerator. Unlike general-purpose hardware such as GPUs, the proposed accelerator integrates specialized rotation and NPU units. This design emphasizes the importance of LightRot in reducing rotation costs, particularly in hardware with dedicated processing units.

In GPUs, rotation operations are handled as part of the general datapath, where both rotation and neural network computations utilize similar hardware resources. As a result, the efficiency gap between these operations is relatively small, and rotation overheads are less critical. However, in dedicated accelerators, the disparity in hardware costs between floating-point (FP) logic for the rotation unit and integer multiply-accumulate (INT MAC) logic for the NPU becomes much more significant, as discussed in Section II-C. This highlights the necessity of minimizing FP overheads in rotation units to maintain overall efficiency, making LightRot's optimizations especially impactful in dedicated hardware designs. The proposed hardware implementation thus serves as a focused demonstration of LightRot's ability to reduce computational and energy costs in specialized accelerator environments.

\begin{figure*}[]
\includegraphics[width=5.0in]{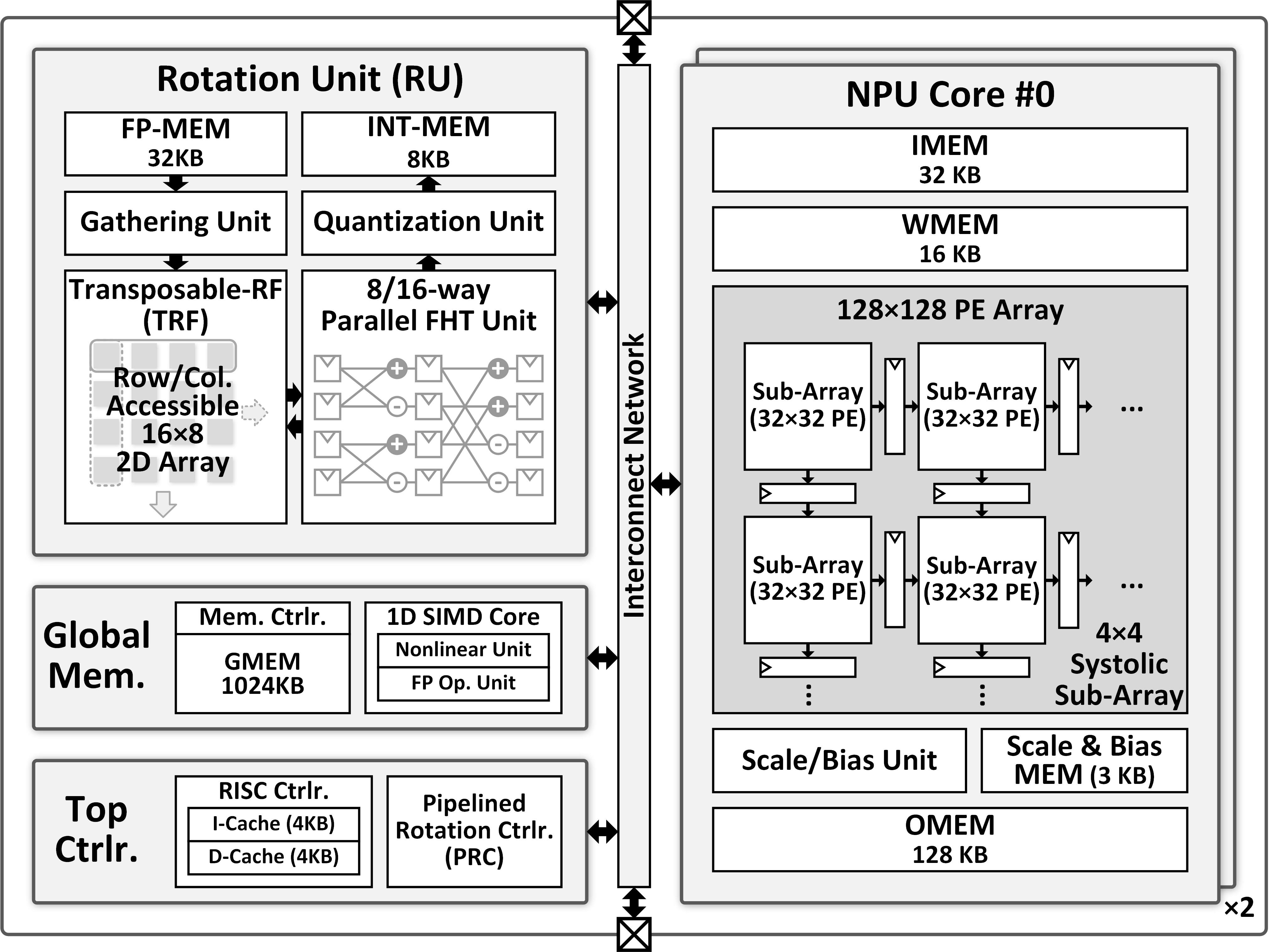}
\centering
\caption{Overall Architecture.}
\label{fig_2_9}
\vspace{-0.2cm}
\end{figure*}

\subsection{Overall Architecture}
The proposed accelerator is designed to maximize the efficiency of low-bit LLM inference by integrating the LightRot algorithm into a dedicated hardware architecture. As illustrated in Fig. 10, the accelerator consists of two main components: the NPU Core, which performs high-throughput integer-based computations for quantized neural networks, and the Rotation Unit (RU), which is optimized for floating-point operations required for rotation tasks. These components are interconnected by a high-bandwidth interconnect network. The Global Memory handles intermediate data storage for activations and weights, while the Top Controller manages pipeline scheduling and synchronization between the NPU Core and the Rotation Unit.

The NPU Core is a highly optimized integer-based processing unit designed for low-bit computations. It features a 128×128 processing element (PE) array, organized into 16 sub-arrays of 32×32 PEs. Within each 32×32 PE sub-array, a weight-stationary datapath is employed, where input activations (IA) are broadcasted to 32 output channels, and partial sums are accumulated through a 32-way reduction tree. Between the 4×4 sub-arrays, a systolic datapath structure is used, operating in a pipelined manner. This hybrid design allows local operations to run in parallel while global operations follow a systolic flow, minimizing the pipeline register count required for systolic datapath operations and ensuring the critical path remains manageable.

Each NPU Core is equipped with 32KB of input activation memory (IMEM) and 16KB of weight memory (WMEM) to store intermediate data for low-bit matrix-vector computations. Additionally, a scale and bias unit facilitates group quantization (GQ) by efficiently applying scale factors and biases during arithmetic operations. These optimizations collectively ensure high throughput for LLM inference while maintaining low energy consumption.

The Rotation Unit (RU) is dedicated to performing online rotation operations, a critical component of the LightRot algorithm. FP memory stores input activations (IA) before rotation, and INT memory holds the quantized IA after rotation. The data flow within the RU begins with FP memory, where IA is read and passed to the Gathering Unit. The Gathering Unit performs permutation operations required for ODA, reordering data as needed to optimize the subsequent rotation. The reordered data is then processed by a Transposable Register File (TRF) and an 8/16-way parallel Fast Hadamard Transform (FHT) Unit, which executes the hierarchical rotation operation with minimal energy and area overhead. After rotation, the Quantization Unit quantizes the rotated IA and stores it in INT memory, making it ready for use in the NPU Core. This streamlined data flow ensures that rotation operations are performed efficiently, with optimized support for GLR and ODA, while minimizing overhead for subsequent low-bit computations.

The Global Memory (GMEM) and Top Controller play crucial roles in data management and task scheduling. The GMEM includes a 1MB memory bank dedicated to storing intermediate activations and weights, ensuring smooth data transfer between components. The Top Controller utilizes a RISC-based architecture to manage pipeline execution and rotation scheduling. To further optimize coordination, the pipelined rotation controller (PRC) ensures synchronization between the NPU Core and the Rotation Unit (RU), effectively hiding rotation latency and maintaining seamless operation across computational stages.

This modular architecture fully capitalizes on the advantages of the LightRot algorithm, with each component optimized for its specific function in low-bit LLM inference. By combining an NPU Core tailored for integer-based neural network computations with a specialized Rotation Unit for efficient floating-point operations, the proposed accelerator delivers high computational efficiency while effectively addressing the challenges posed by rotation overhead.

\subsection{Rotation Unit with Hierarchical FHT Operation}

\begin{figure}[]
\centering
\subfigure[]{
\includegraphics[width=3.4in]{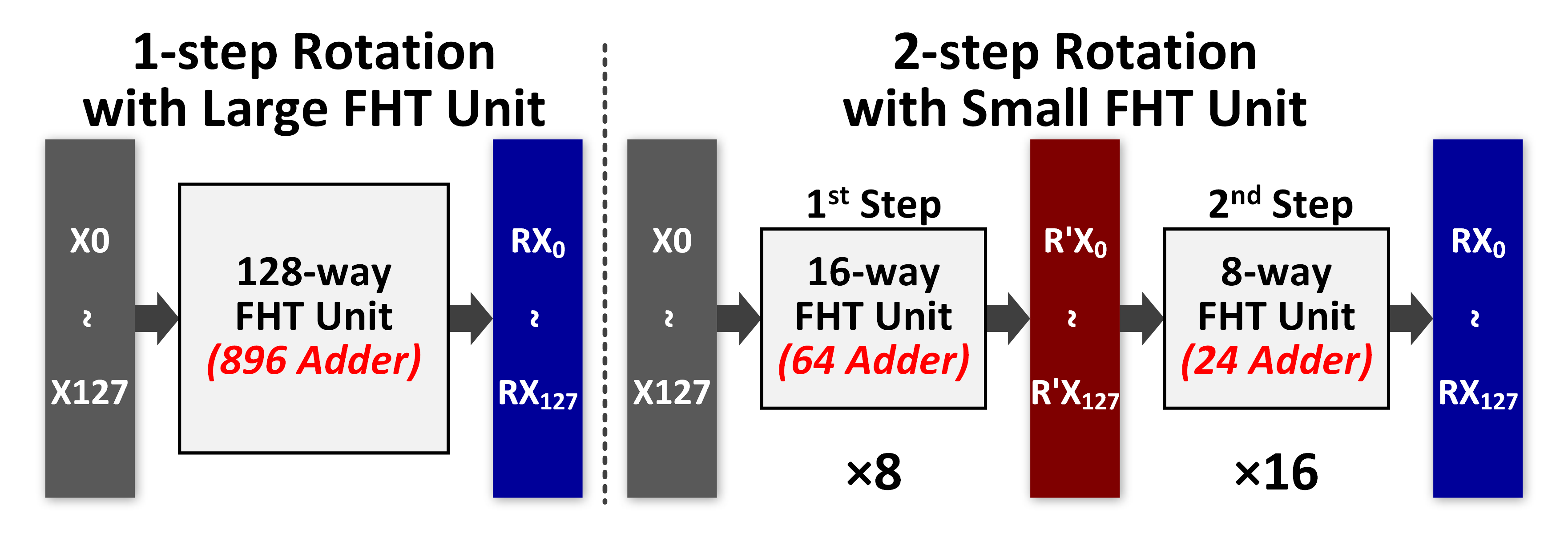}
\centering
}
\subfigure[]{
\includegraphics[width=3.4in]{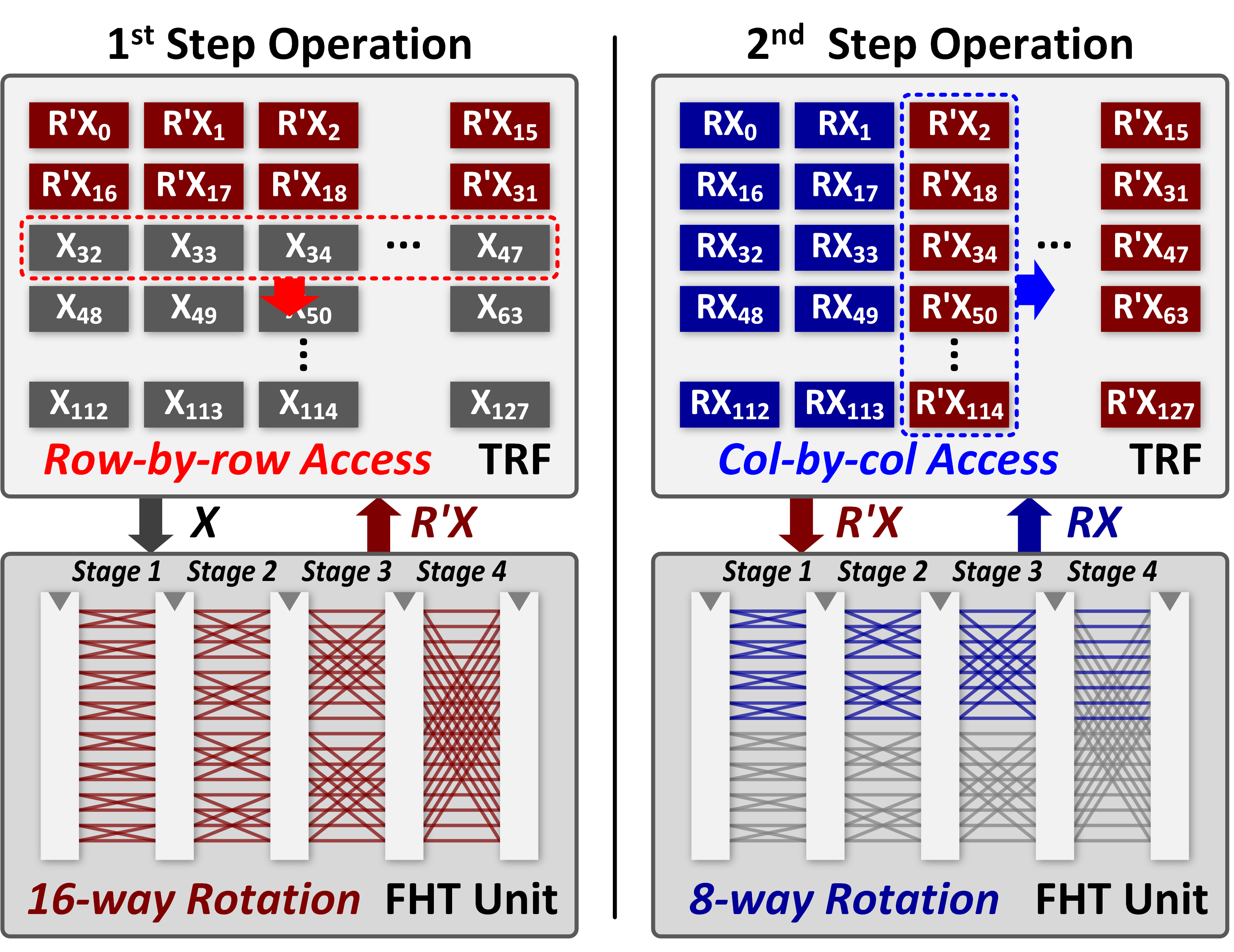}
\centering
}
\subfigure[]{
\includegraphics[width=3.4in]{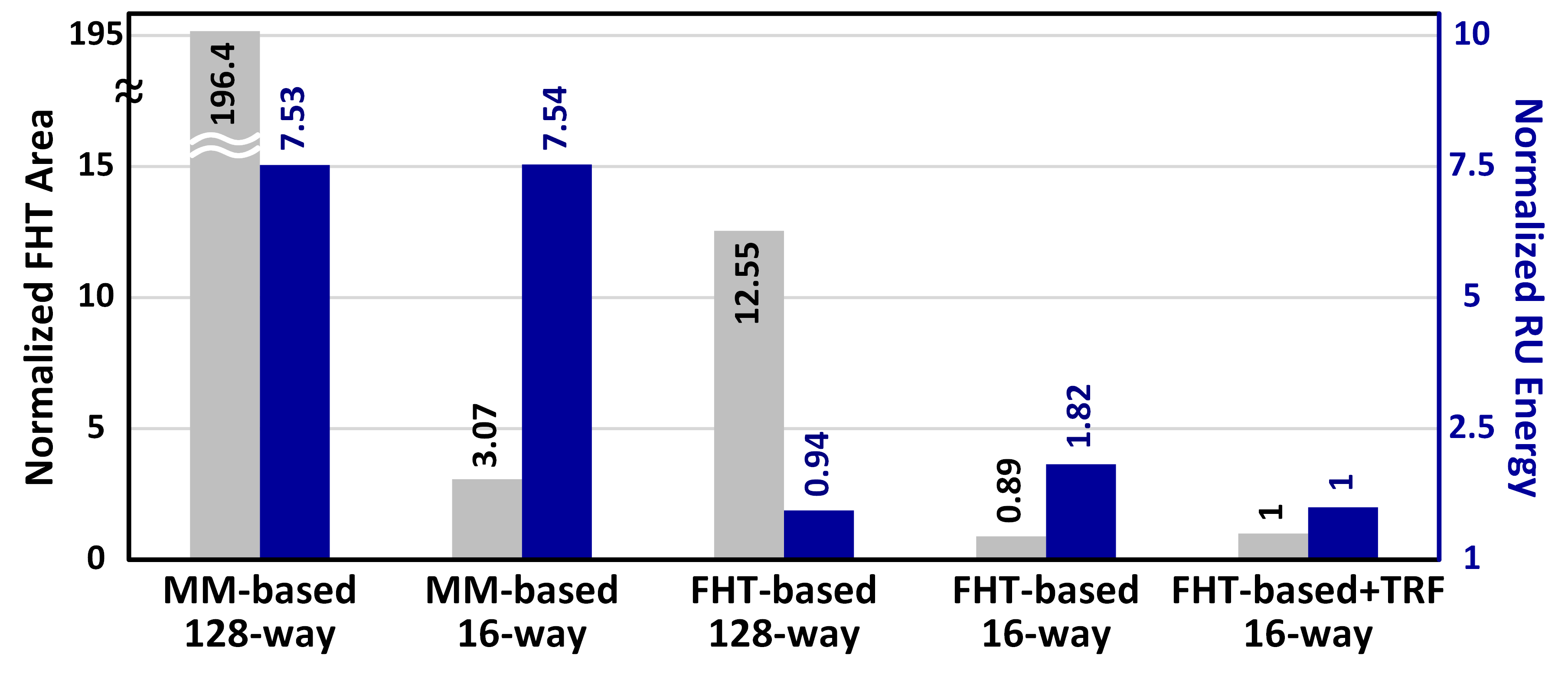}
\centering
}
\centering
\caption{Hierarchical FHT Operation with Transposable-RF (TRF) for GLR.}
\vspace{-6pt}
\end{figure}

The Rotation Unit (RU) leverages a hierarchical Fast Hadamard Transform (FHT) operation to efficiently handle large-scale rotation tasks while minimizing area and energy overhead. As shown in Fig. 11(a), the traditional one-step FHT operation requires a large FHT unit, such as a 128-way FHT, which introduces significant hardware complexity (e.g., 896 adders). To address this, the proposed design employs a two-step hierarchical FHT approach, breaking the rotation operation into smaller units. In the 2-stage FHT unit, the size of the small FHT unit was determined to match the target group size of 128-way. For ease of hardware implementation, a power-of-two number should be used. Among them, the smallest number 16, whose square value is greater than 128 was selected for 2-stage operation.

In the first step, the RU uses a 16-way FHT Unit with 64 adders to perform partial rotations across smaller subgroups. These partial results are then processed in the second step using an 8-way FHT Unit with only 24 adders. By dividing the operation into two stages, the RU significantly reduces hardware requirements compared to a single large FHT unit. This approach achieves a balance between computational efficiency and hardware scalability, as illustrated in the right-hand side of Fig. 11(a).

Fig. 11(b) highlights the data flow for the hierarchical FHT operation, where the 8/16-way parallel FHT Unit and the Transposable Register File (TRF) collaborate to support 2-step rotation within a single rotation task. In the 2-step operation, the first and second steps require data to be loaded with different strides, which is efficiently managed using the 16×8 TRF’s transposable datapath. 
Initially, 128 input activations (IA) are loaded from FP-MEM into the TRF. During the first step, the TRF performs row-by-row access to load 16 consecutive IAs with a stride of 1. These values are then passed to the 16-way FHT Unit for processing, and the resulting FHT outputs are updated back into the corresponding rows of the TRF.
In the second step, the TRF switches to column-by-column access with a stride of 16, loading 8 IAs at a time. These values are processed by the 8-way FHT Unit, where half of the 16-way FHT Unit is gated to operate in 8-way mode, saving energy and hardware resources. The resulting outputs from this step are written back to the respective columns of the TRF.
This hierarchical structure minimizes data movement, reduces memory bandwidth requirements, and ensures high throughput, making it an effective solution for hardware-efficient rotation tasks.

The benefits of this design are further quantified in Fig. 11(c), where the hierarchical FHT operation (FHT-based + TRF) achieves significant area and energy savings compared to both matrix multiplication (MM)-based and single-step FHT-based approaches. Specifically, the proposed design reduces FHT logic area by 12.55 times compared to the 128-way FHT-based rotation unit and energy by 1.82 times compared to the 16-way FHT-based rotation unit without TRF. These results demonstrate that the hierarchical FHT operation effectively balances performance and hardware efficiency, making it well-suited for large-scale LLM inference tasks.

\subsection{Gathering Unit for ODA}
The Gathering Unit is designed to enable the reordering of input activations (IA) during the loading process from FP-MEM to the TRF, supporting the proposed ODA technique. This unit ensures that outliers are identified and reordered before being stored in the TRF, optimizing data distribution for high quantizability with ODA.

As shown in Fig. 13, the Gathering Unit consists of key components. The Channel Counter iterates through the 128 channels of IA to generate initial addresses. Simultaneously, the Group ID Counter ensures that outliers are loaded as the first entry in each group to facilitate ODA operations. The Outlier Address Generator identifies outlier channels based on precomputed indices stored in the Outlier Channel Table, which are generated offline using calibration datasets. These indices are sorted offline and accessed sequentially by the Group ID Counter and Table Counter. The Inlier Address Generator handles the sequential generation of inlier addresses by filtering out addresses that correspond to outlier channels.

The data flow begins with the Channel Counter, which checks whether the current channel is the first in the group. For the first channel in a group, the Group ID Counter directs the Outlier Address Generator to fetch the data corresponding to the outlier address assigned to the current group and loads it from FP-MEM to the TRF. For all other channels, the Inlier Address Generator sequentially generates inlier addresses. To avoid outlier addresses, the Address Counter increments sequentially while the Table Counter maintains the next outlier address in the Outlier Buffer. Initially, the Table Counter fetches the first outlier address and stores it in the Outlier Buffer. When the Address Counter matches the value in the Outlier Buffer, it skips that address (since it corresponds to an outlier) and increments the Table Counter to fetch the next outlier address into the Outlier Buffer. By keeping the outlier addresses sorted in the Outlier Channel Table, the Outlier Buffer always holds the nearest future outlier address. This mechanism ensures that the Inlier Address Generator produces only inlier addresses, avoiding conflicts with outliers.

This approach ensures that the ODA algorithm can effectively align outliers to the all-1 row of the rotation matrix, as discussed in Section III-C
. By preprocessing the data at the memory interface, the Gathering Unit minimizes runtime overhead while supporting hierarchical FHT operations in the RU.

\begin{figure}[]
\includegraphics[width=3.4in]{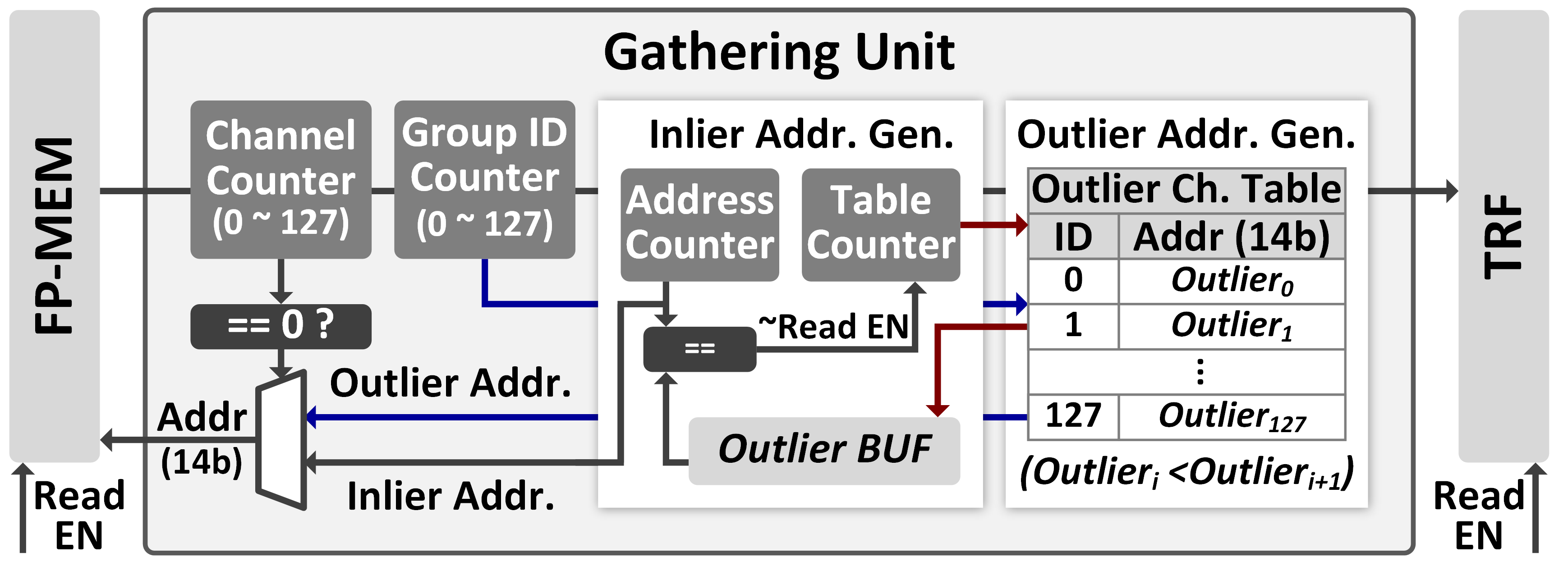}
\centering
\caption{Gathering Unit for ODA.}
\label{fig_2_10}
\vspace{-0.2cm}
\end{figure}

\section{Implementation Result}

\begin{comment}
\begin{figure}[]
\includegraphics[width=3.4in]{Figure/F14.png}
\centering
\caption{Group-Pipelined Rotation (GPR).}
\label{fig_2_11}
\vspace{-0.2cm}
\end{figure}
\section{Implementation Results}
\end{comment}

\begin{table*}[]
\caption{Comparison Table.}
\includegraphics[width=6.0in]{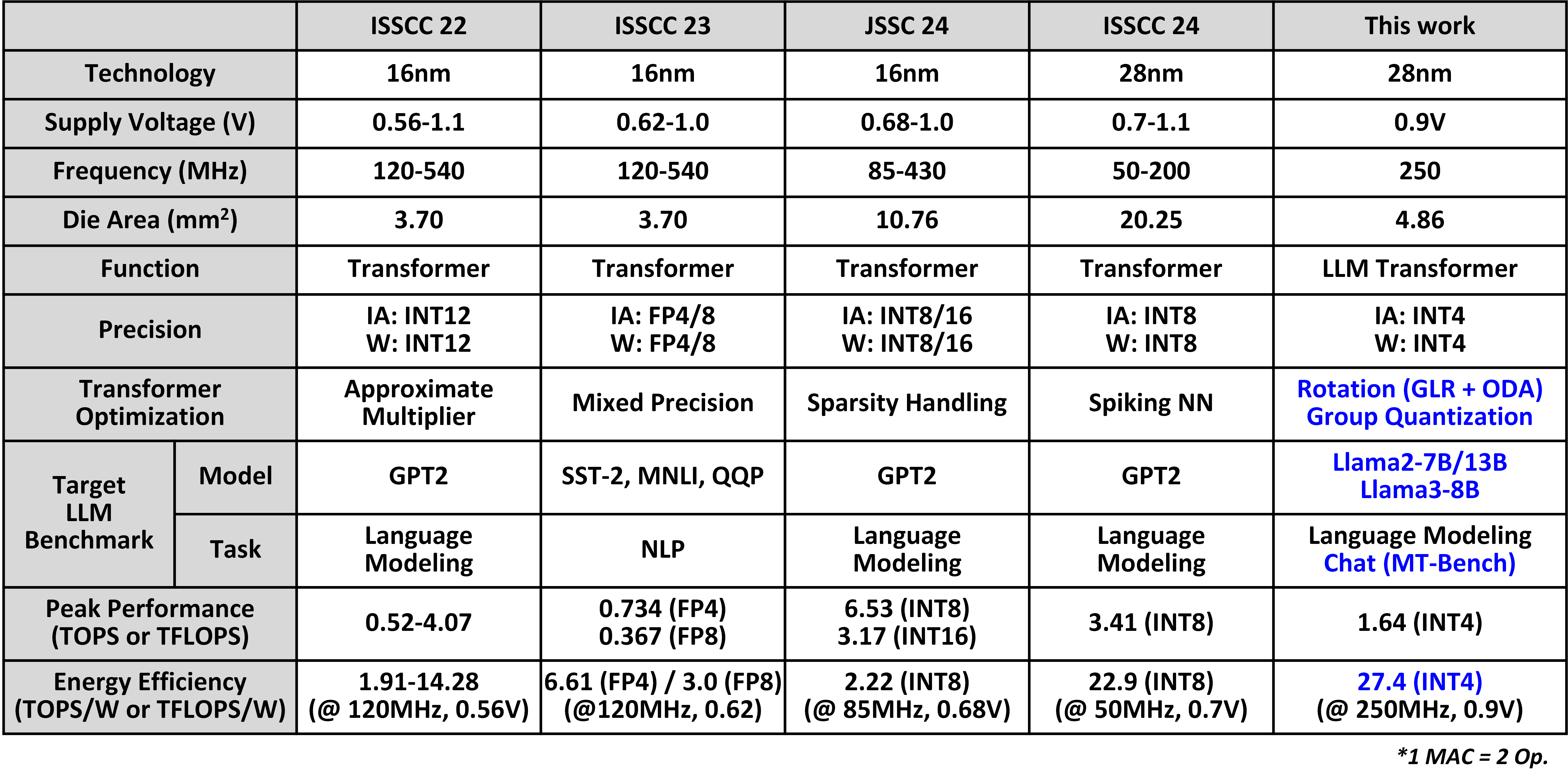}
\centering
\label{table_2}
\end{table*}

\begin{figure}[]
\centering
\subfigure[]{
\includegraphics[width=3.0in]{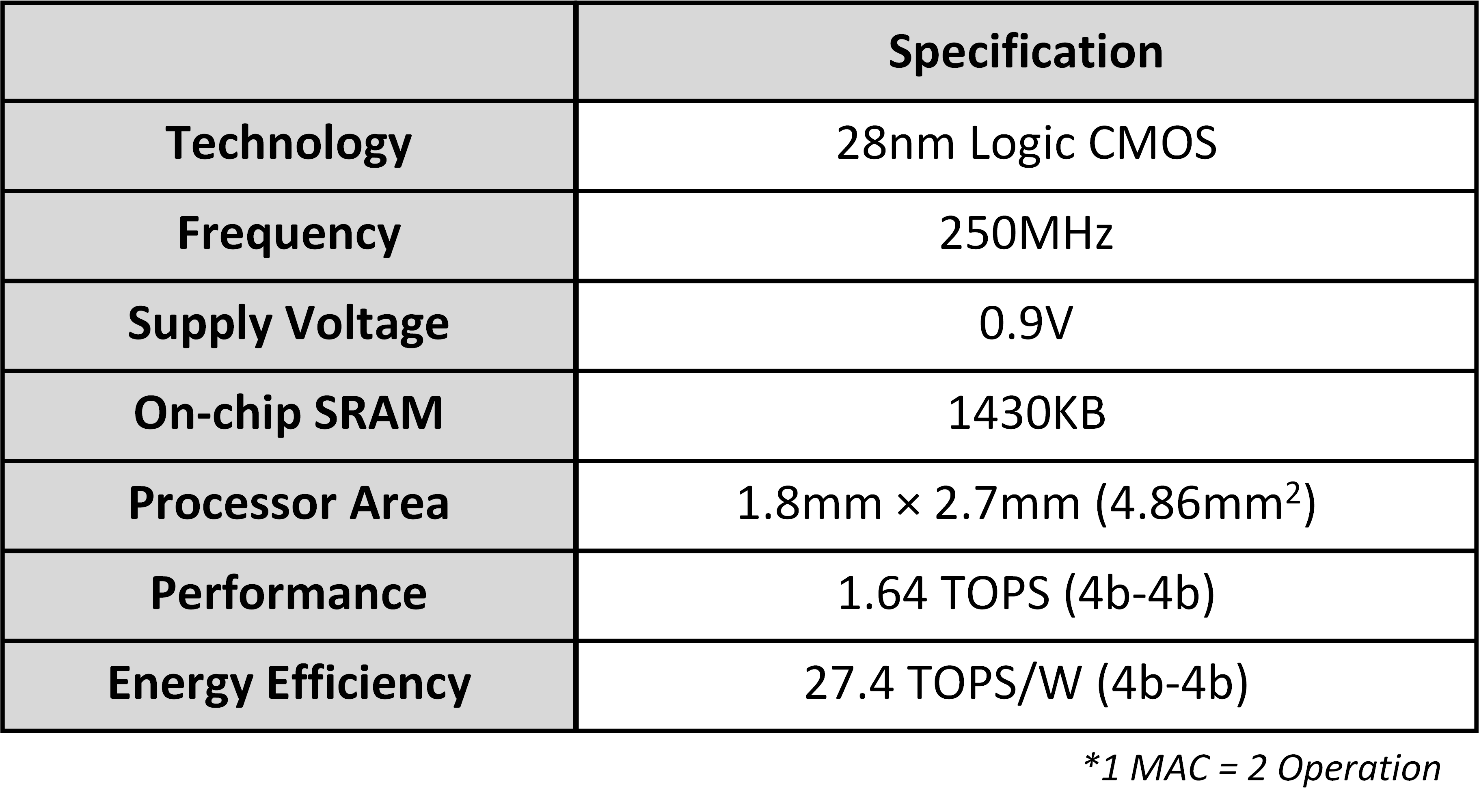}
\centering
}
\subfigure[]{
\includegraphics[width=3.4in]{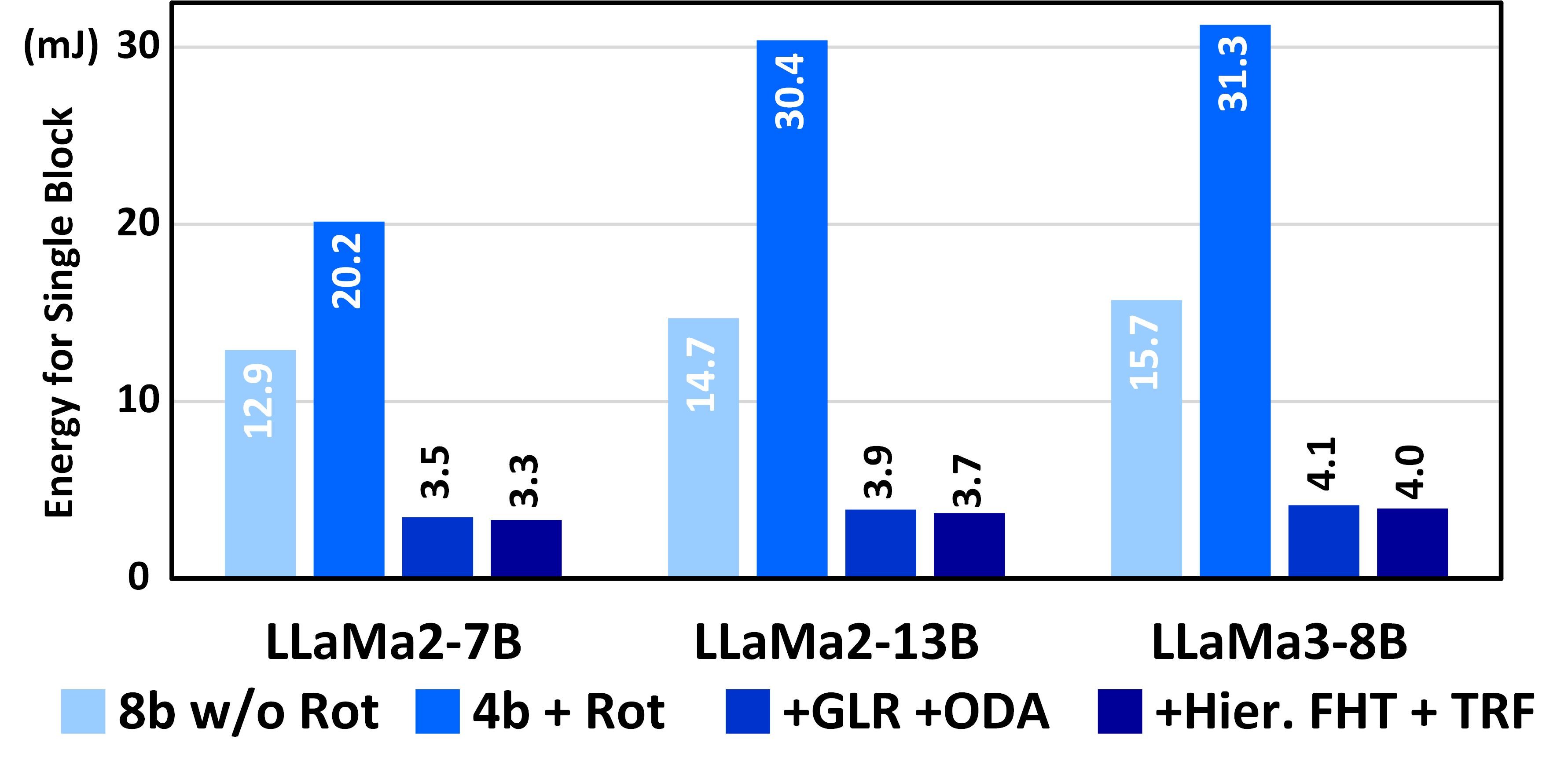}
\centering
}
\subfigure[]{
\includegraphics[width=3.4in]{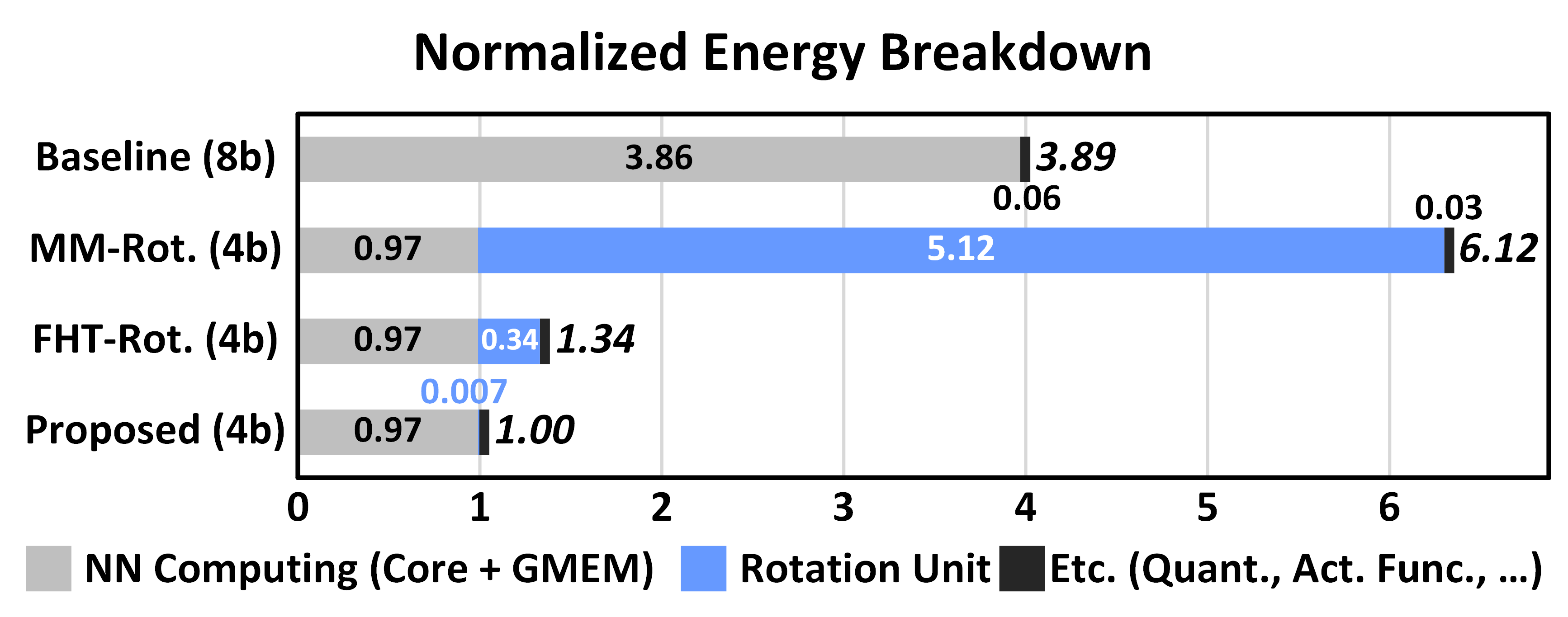}
\centering
}
\centering
\caption{(a) Chip Performance Summary (b) Energy Consumption on LLaMA Prefilling Stage (Input Token Length = 128) (c) Energy Breakdown}
\end{figure}

\begin{figure}[]
\includegraphics[width=2.8in]{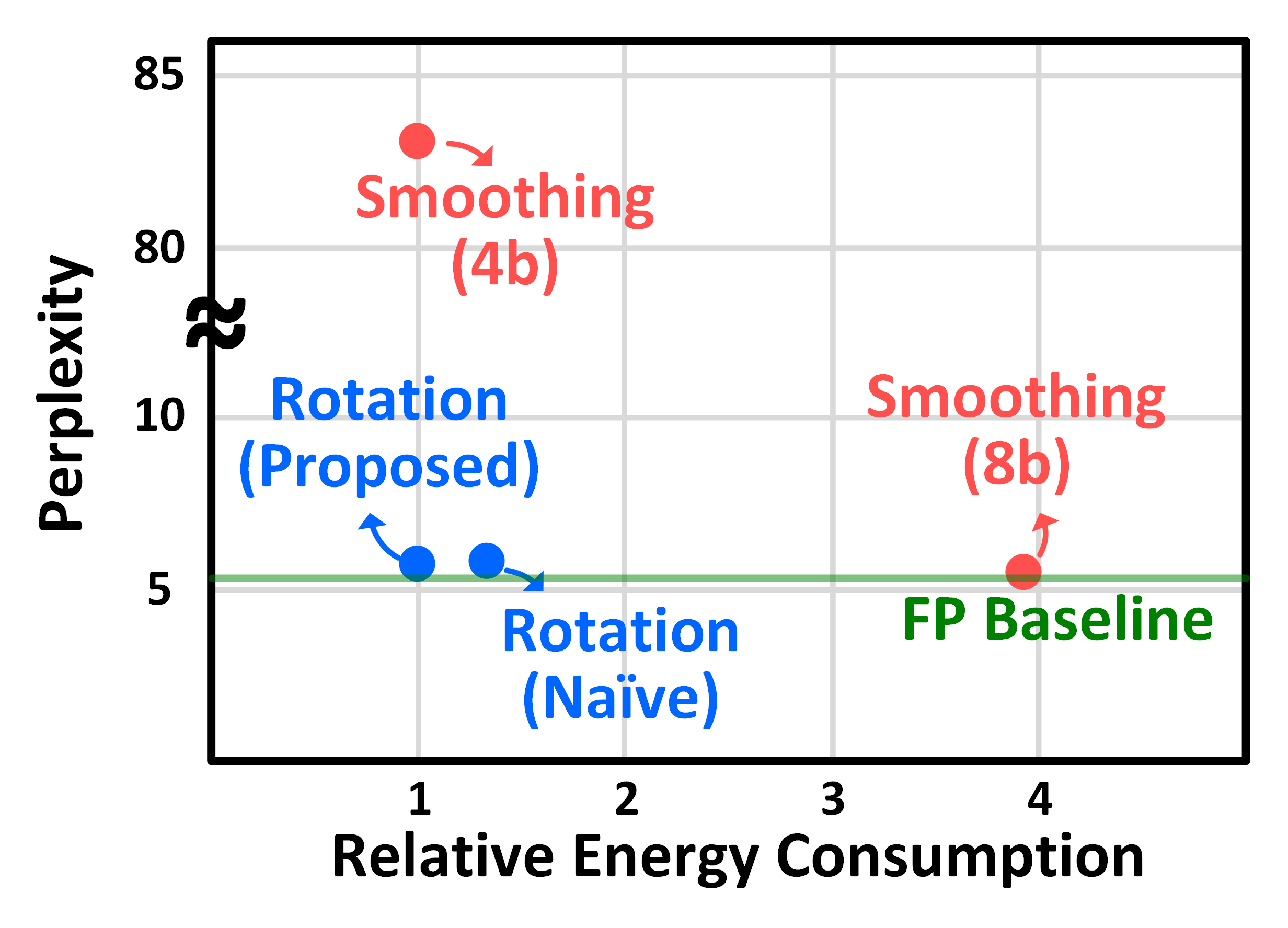}
\centering
\caption{Accuracy-Energy Comparison of Different Quantization Configuration (LLaMMA2-7B)}
\label{fig_2_12}
\vspace{-0.2cm}
\end{figure}

\begin{comment}
\begin{figure}[]
\includegraphics[width=3.4in]{Figure/F17.png}
\centering
\caption{Chip Phptograph.}
\label{fig_2_13}
\vspace{-0.2cm}
\end{figure}
\end{comment}

Fig. 13(a) summarizes the key specifications of the proposed accelerator implemented in a 28nm logic CMOS process. Operating at a frequency of 250 MHz and supply voltage of 0.9V, the accelerator includes 1430 KB of on-chip SRAM and occupies a die area of 4.86 mm². It achieves a peak performance of 1.64 TOPS for 4b-quantized operations and an energy efficiency of 27.4 TOPS/W. These specifications highlight the accelerator’s capability to deliver high computational performance and energy efficiency, specifically tailored for 4b-quantized LLM inference tasks.

Fig. 13(b) illustrates the energy consumption per LLM block across different configurations for the evaluated models: LLaMA2-7B, LLaMA2-13B, and LLaMA3-8B. When comparing 8b w/o Rot to configurations using 4b + Rot, the energy consumption initially increases due to the overhead introduced by rotation operations. For example, straightforwardly applying rotation results in significant power increases, especially for larger models like LLaMA3-8B, where the energy consumption reaches 31.3 mJ per block compared to 15.7 mJ for the 8-bit baseline.

The proposed GLR and ODA techniques, combined with the FHT-based rotation unit, significantly reduce energy consumption. These optimizations minimize the overhead of rotation while maintaining high quantization accuracy. For instance, the inclusion of GLR and ODA, alongside the hierarchical FHT structure, achieves a dramatic reduction in energy usage. For the LLaMA3-8B model, the energy consumption drops to 4.0 mJ, representing a 3.9× improvement compared to the baseline configuration (8b w/o Rot). The combination of GLR and ODA enables efficient management of rotation costs, ensuring substantial energy savings while supporting 4-bit quantized inference.

Fig. 13(c) illustrates the energy breakdown before and after applying rotation optimizations. While neural network computing energy is significantly reduced through rotation-based quantization, MM-based rotation introduces substantial energy overhead, making it inefficient. The adoption of FHT-based rotation mitigates this issue to some extent, but it still results in noticeable overhead. The proposed LightRot approach further optimizes the rotation unit, effectively minimizing the associated energy cost to a negligible level, demonstrating its superiority in enabling energy-efficient LLM inference. Additionally, while on-the-fly activation quantization requires floating-point division, this can be efficiently replaced with comparison-based operations as proposed in \cite{edgediff}, resulting in an energy overhead of less than 3\%.

Fig. 14 presents the accuracy-energy comparison of different quantization configurations on the LLaMA2-7B model. In smoothing-based quantization \cite{smoothquant}, the 4-bit configuration achieves low energy consumption but suffers from a significant perplexity drop. Conversely, the 8-bit smoothing-based quantization maintains higher accuracy but incurs substantial energy overhead. In contrast, rotation-based quantization preserves FP-level perplexity even at 4-bit precision while significantly reducing energy consumption. Although naive rotation initially introduces energy overhead compared to smoothing-based 4-bit quantization, the proposed LightRot approach effectively mitigates this overhead, achieving both high accuracy and energy efficiency.

Table II provides a comparative analysis of the proposed accelerator against previous state-of-the-art implementations \cite{isscc2022, isscc2023, ayaka, ctransformer}. Existing processors relied heavily on optimization techniques that are not tailored for low-bit LLM inference. As a result, these designs typically employed higher precision operations (e.g., INT8/INT12), which inherently limited their energy efficiency, especially for power-critical applications. 

Several existing accelerators have also been optimized for INT4 \cite{edgediff, vsq} or even lower-precision inference \cite{arbi}. However, these approaches have primarily been validated on discriminative models, with limited verification on LLMs, which require significantly higher accuracy. Furthermore, to support low-bit inference while avoiding excessive quantization errors, many of these designs rely on mixed-precision \cite{edgediff, vsq} or arbitrary quantization \cite{arbi}. While these techniques help mitigate accuracy loss, they also introduce substantial hardware complexity, making them less efficient for dedicated LLM acceleration.

These limitations highlight the significance of LightRot’s approach, which is specifically optimized for low-bit LLM inference while maintaining high accuracy. Unlike previous accelerators that either relied on higher-precision quantization or mixed-precision designs to mitigate quantization errors, LightRot fully exploits rotation-based quantization to enable efficient INT4 inference without requiring excessive precision adjustments. By integrating GLR and ODA, the proposed method effectively addresses outlier issues and maintains competitive accuracy, as demonstrated in MT-Bench evaluations. This specialized approach not only reduces the energy overhead of rotation but also enhances the quantizability of the data, enabling state-of-the-art energy efficiency. The proposed design achieves 27.4 TOPS/W for INT4 operations, significantly outperforming previous works, demonstrating efficiencies ranging from 1.91–22.9 TOPS/W.

Moreover, most previous processors primarily evaluated their performance on benchmarks such as GPT2 for language modeling tasks. These benchmarks fail to capture the complexity and real-world requirements of conversational applications. In contrast, the proposed accelerator has been rigorously evaluated using MT-Bench, a benchmark designed for chat scenarios. This evaluation demonstrates the effectiveness of the proposed architecture in real-world applications such as conversation and chat tasks, further validating its practicality for modern LLM-based systems.

By addressing the limitations of existing designs and focusing on real-world applicability, this work sets a new standard for energy-efficient low-bit LLM inference, offering a comprehensive solution for both language modeling and conversational AI tasks.

\section{Discussion}
\subsection{Comparison with Microscaling}

Recently, the industry has introduced microscaling (MX) \cite{microscaling} as a data format to support low-bit LLM inference. Similar to traditional group quantization, MX applies a per-vector scaling factor, but it uses a relatively small vector size (32) while minimizing hardware costs with the scaling factor in an exponent-only format (E8M0). This characteristic allows MX to effectively cover a wide range of data values, making it particularly advantageous for training rather than inference.

As noted in \cite{amxfp4}, the effectiveness of MX in handling outliers during inference is inferior to rotation-based methods, resulting in lower accuracy. To address this limitation, \cite{amxfp4} proposed combining MX with asymmetric quantization. However, this approach introduces additional overhead for zero-point handling and requires a higher-precision scaling factor, reducing compatibility with the existing MX standard hardware.

Similar to \cite{amxfp4}, the proposed LightRot also employs asymmetric quantization due to the application of ODA. However, LightRot utilizes a larger group size (128), which significantly reduces the overhead associated with zero-point and scaling factor handling. This makes LightRot a more efficient solution for low-bit LLM inference while maintaining compatibility with practical hardware constraints.

\section{Conclusion}
This work presents LightRot, a novel lightweight rotation scheme and dedicated hardware architecture for accurate low-bit inference of large language models (LLMs). By integrating the proposed GLR and ODA algorithms with a hierarchical FHT-based rotation unit, the architecture addresses critical challenges in low-bit quantization, including outlier handling and the energy overhead of rotation operations. Through rigorous evaluations, the proposed accelerator achieves a state-of-the-art energy efficiency of 27.4 TOPS/W, surpassing prior works that rely on higher precision inference methods. Unlike existing accelerators focusing on traditional language modeling benchmarks such as GPT2, this work demonstrates its practicality on advanced LLMs like LLaMA2 and LLaMA3. By leveraging MT-Bench, the proposed design also validates its effectiveness in real-world conversational scenarios, showcasing its applicability to modern AI tasks. This work paves the way for more energy-efficient and scalable hardware solutions tailored to the demands of next-generation LLM applications.

\vspace{11pt}
\begin{IEEEbiography}[{\includegraphics[width=1in,height=1.25in,clip,keepaspectratio]{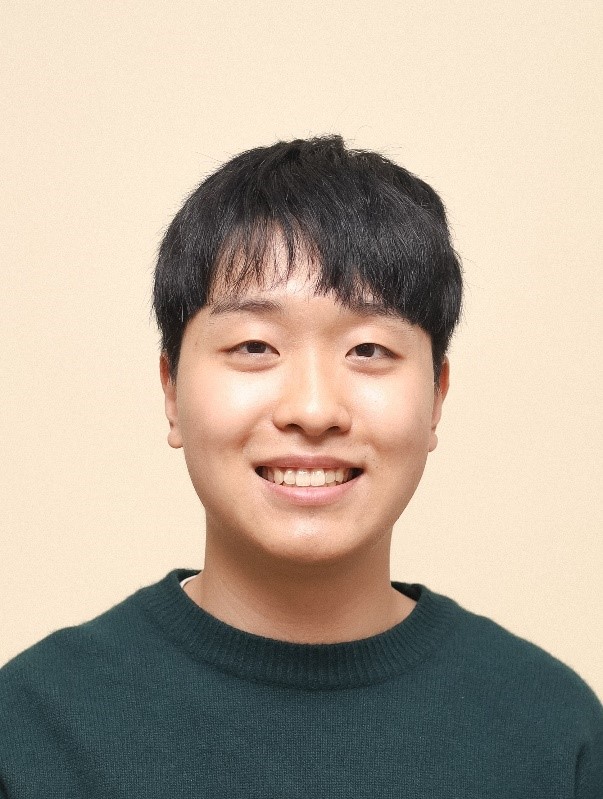}}]{Sangjin Kim}
(Member, IEEE) received his B.S., M.S., and Ph.D. degrees in Electrical Engineering from the Korea Advanced Institute of Science and Technology (KAIST), Daejeon, South Korea, in 2019, 2021, and 2024, respectively. He is currently a Postdoctoral Associate at KAIST. 

His research interests include computing-in-memory for low-power AI accelerators, processing-in-memory for energy-efficient AI systems, and hardware-software co-optimization for generative AI models.
\end{IEEEbiography}

\begin{IEEEbiography}[{\includegraphics[width=1in,height=1.25in,clip,keepaspectratio]{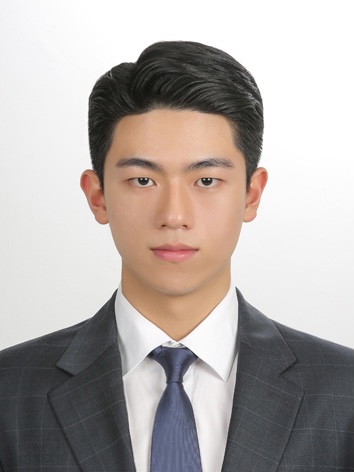}}]{Yuseon Choi}
(Graduate Student Member, IEEE) received his B.S. degree in Electrical Engineering from Korea Advanced Institute of Science and Technology (KAIST), Daejeon, South Korea, in 2024, where he is currently pursuing the M.S. degree. His current research interests include HW-SW co-optimization for generative AI models, deep learning accelerators for robotics, and energy-efficient system-on-chip designs.
\end{IEEEbiography}

\begin{IEEEbiography}[{\includegraphics[width=1in,height=1.25in,clip,keepaspectratio]{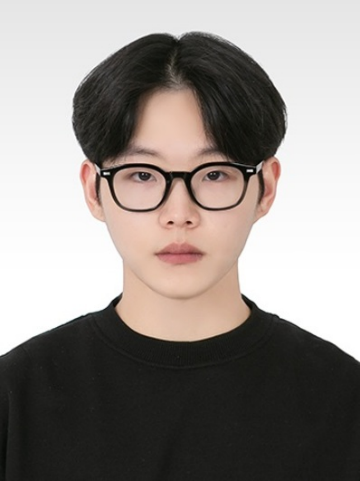}}]{Jungjun Oh}
(Graduate Student Member, IEEE) received the B.S. in electronic engineering from Sungkyunkwan University, Seoul, South Korea, in 2024. He is currently pursuing an M.S. degree in Graduate School of AI Semiconductor, Korea Advanced Institute of Science and Technology (KAIST), Daejeon, South Korea. His current research interests include hardware-aware optimization algorithms and energy-efficient deep learning processors.
\end{IEEEbiography}

\begin{IEEEbiography}[{\includegraphics[width=1in,height=1.25in,clip,keepaspectratio]{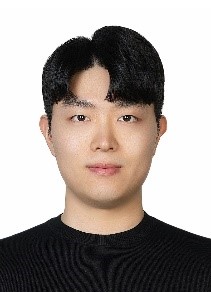}}]{Byeongcheol Kim}
(Graduate Student Member, IEEE) received the B.S. in electronic engineering from Sogang University, Seoul, South Korea, in 2018. He is currently pursuing the M.S. degree in Graduate School of AI Semiconductor, Korea Advanced Institute of Science and Technology (KAIST), Daejeon, South Korea. His current research interests include energy-efficient processing-in-memory accelerators and deep learning processors.
\end{IEEEbiography}

\begin{IEEEbiography}[{\includegraphics[width=1in,height=1.25in,clip,keepaspectratio]{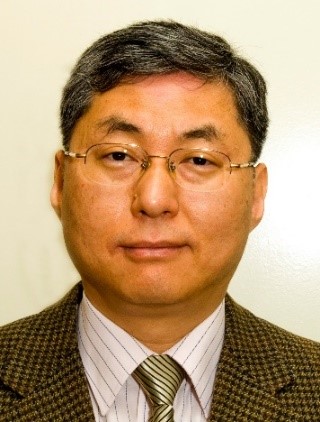}}]{Hoi-Jun Yoo}
(Fellow, IEEE) graduated from the Department of Electronics, Seoul National University, Seoul, South Korea, in 1983. He received the M.S. and Ph.D. degrees in electrical engineering from the Korea Advanced Institute of Science and Technology (KAIST), Daejeon, South Korea, in 1985 and 1988, respectively. 

Dr. Yoo has served as a member of the Executive Committee for the International Solid-State Circuits Conference (ISSCC), the Symposium on Very Large-Scale Integration (VLSI), and the Asian Solid-State Circuits Conference (A-SSCC), the TPC Chair of the A-SSCC 2008 and the International Symposium on Wearable Computer (ISWC) 2010, the IEEE Distinguished Lecturer from 2010 to 2011, the Far East Chair of the ISSCC from 2011 to 2012, the Technology Direction Sub-Committee Chair of the ISSCC in 2013, the TPC Vice-Chair of the ISSCC in 2014, and the TPC Chair of the ISSCC in 2015. More details are available at http://ssl.kaist.ac.kr
\end{IEEEbiography}
\vfill

\end{document}